\documentclass[aps, prx,amsmath,amssymb,longbibliography,showpacs,letterpaper,twocolumn,nofootinbib,floatfix,superscriptaddress]{revtex4-2}

\usepackage{tikz}
\usepackage{amsmath}
\usepackage{graphicx}
\usepackage{fullpage}
\usepackage[colorlinks=true,allcolors=blue]{hyperref}
\usepackage[capitalise]{cleveref}
\usepackage[utf8]{inputenc}
\usepackage{siunitx}
\usepackage{url}
\usepackage{rotating}
\usepackage{epsfig,graphics,rotate}
\usepackage{wrapfig}
\usepackage[caption=false]{subfig}
\usepackage{bm}
\usepackage{amssymb}
\usepackage{amsmath}
\usepackage{amsfonts}
\usepackage{booktabs}
\usepackage{microtype}
\usepackage{multirow, makecell}
\usepackage{lipsum}
\usepackage{xspace}
\usepackage{comment}
\usepackage{floatrow}
\usepackage{ragged2e}

\usepackage{nicefrac, xfrac}
\usepackage{mathtools} 
\usepackage{relsize}   

\usepackage{float}

\usepackage{psvectorian}
\usepackage{bbding}

\DeclareSIUnit \s {\second}
\DeclareSIUnit \ns {\nano\second}
\DeclareSIUnit \mus {\micro\second}
\DeclareSIUnit \ms {\milli\second}
\DeclareSIUnit \MB {\mega\byte}
\DeclareSIUnit \GB {\giga\byte}
\DeclareSIUnit \TB {\tera\byte}
\DeclareSIUnit \PB {\peta\byte}
\DeclareSIUnit \Mbps {\mega\bit/\s}
\DeclareSIUnit \Gbps {\giga\bit/\s}
\DeclareSIUnit \Tbps {\tera\bit/\s}
\DeclareSIUnit \Pbps {\peta\bit/\s}
\DeclareSIUnit \kton {\kilo\tonne} 
\DeclareSIUnit \kt {\kilo\tonne}
\DeclareSIUnit \Mt {\mega\tonne}
\DeclareSIUnit \eV {\electronvolt}
\DeclareSIUnit \keV {\kilo\electronvolt}
\DeclareSIUnit \MeV {\mega\electronvolt}
\DeclareSIUnit \GeV {\giga\electronvolt}
\DeclareSIUnit \TeV {\tera\electronvolt}
\DeclareSIUnit \PeV {\peta\electronvolt}
\DeclareSIUnit \EeV {\exa\electronvolt}
\DeclareSIUnit \m {\meter}
\DeclareSIUnit \cm {\centi\meter}
\DeclareSIUnit \in {\inchcommand}
\DeclareSIUnit \km {\kilo\meter}
\DeclareSIUnit \kV {\kilo\volt}
\DeclareSIUnit \kW {\kilo\watt}
\DeclareSIUnit \MW {\mega\watt}
\DeclareSIUnit \MHz {\mega\hertz}
\DeclareSIUnit \mrad {\milli\radian}
\DeclareSIUnit \year {years}
\DeclareSIUnit \POT {POT}
\DeclareSIUnit \sig {$\sigma$}
\DeclareSIUnit\parsec{pc}
\DeclareSIUnit\lightyear{ly}
\DeclareSIUnit\foot{ft}
\DeclareSIUnit\ft{ft}
\DeclareSIUnit \ppb{ppb}
\DeclareSIUnit \ppt{ppt}
\DeclareSIUnit \samples{S}
\DeclareSIUnit \pe{PE}
\DeclareSIUnit \kpc{kpc}

\definecolor{lime}{HTML}{A6CE39}
\DeclareRobustCommand{\orcidicon}{
	\begin{tikzpicture}
	\draw[lime, fill=lime] (0,0) 
	circle [radius=0.16] 
	node[white] {{\fontfamily{qag}\selectfont \tiny ID}};
	\draw[white, fill=white] (-0.0665,0.095) 
	circle [radius=0.005];
	\end{tikzpicture}
	\hspace{-2mm}
}

\usepackage{booktabs}
\usepackage{siunitx}
\usepackage{tabularx}

\foreach \x in {A, ..., Z}{\expandafter\xdef\csname orcid\x\endcsname{\noexpand\href{https://orcid.org/\csname orcidauthor\x\endcsname}
			{\noexpand\orcidicon}}
   }

\begin{document}

\author{Miller~MacDonald}
\email{mmacdonald@college.harvard.edu}
\affiliation{Department of Physics \& Laboratory for Particle Physics and Cosmology, Harvard University, Cambridge, MA 02138, USA}

\author{Kiara~Carloni}
\email{kcarloni@g.harvard.edu}
\affiliation{Department of Physics \& Laboratory for Particle Physics and Cosmology, Harvard University, Cambridge, MA 02138, USA}

\author{Carlos~A.~Arg{\"u}elles}
\email{carguelles@fas.harvard.edu}
\affiliation{Department of Physics \& Laboratory for Particle Physics and Cosmology, Harvard University, Cambridge, MA 02138, USA}

\author{Ivan~Mart\'inez-Soler}
\email{ivan.j.martinez-soler@durham.ac.uk}
\affiliation{Institute for Particle Physics Phenomenology, Durham University, South Road, DH1 3LE, Durham, UK}

\author{Rafael~\surname{Alves Batista}}
\email{rafael.alves\_batista@iap.fr}
\affiliation{Sorbonne Universit\'e, Institut d'Astrophysique de Paris, CNRS, UMR 7095, 98 bis bd Arago, 75014 Paris, France}

\title{Exploring New Propagation Scales With Galactic Neutrinos}
\begin{abstract}
{
The recent observation of high-energy Galactic neutrinos by IceCube allows for searches of new physics affecting neutrino propagation on scales of $\mathcal{O}(10^9-10^{15})\,\mathrm{km/GeV}$ in distance over energy.
We assess the sensitivity of upcoming measurements of Galactic neutrinos by IceCube and KM3NeT to such new phenomena.
We focus on two scenarios: quasi-Dirac neutrinos and neutrino decays.
In the quasi-Dirac scenario, we find that joint measurements by IceCube and KM3NeT are sensitive to the mass-squared differences $\delta m^2 \in \left(10^{-13.6}~\mathrm{eV^2}, 10^{-12.3}~\mathrm{eV^2}\right)$ at the $90\%$ confidence level. 
For neutrino decays, the same measurements are sensitive to mass over lifetime ratios $m / \tau > 10^{-12.8}~\mathrm{eV^2}$ at the same significance. 
Our results demonstrate that measurements of Galactic neutrinos by a global network of neutrino telescopes can probe signatures of neutrino mass models.}
\end{abstract}

\maketitle

\section{Introduction}

The discovery of neutrino oscillations through measurements of solar~\cite{Super-Kamiokande:2001ljr,SNO:2002tuh,Borexino:2008dzn} and atmospheric~\cite{Super-Kamiokande:1998kpq,IceCube:2014flw,KM3NeT:2024ecf} neutrinos implies that neutrinos have a non-zero mass. The presence of massive neutrinos requires extensions to the Standard Model (SM). 
However, the scale associated with the mass generation mechanism is unknown.
Thus, the discovery of neutrino oscillations was largely an accident of distance and energy scales.
It was an accident that the atmospheric neutrino beam and Earth scale yielded sensitivity to $\Delta m^2 \sim \mathcal{O}(10^{-3})\,\mathrm{eV^2}$, and it was also an accident that the conditions of the solar interior and neutrino energies produced significant disappearance of electron neutrinos. 
Without a proven theory, other scales associated with the mystery of neutrino masses are only guesses.

One possible extension to the SM that produces effects on new lengthscales is the quasi-Dirac neutrino scenario~\cite{Wolfenstein:1981kw,Petcov:1982ya,Valle:1983dk,Kobayashi:2000md}. 
If neutrinos are quasi-Dirac (QD), then we expect new oscillation scales in the neutrino spectrum parameterized by the hyperfine mass splitting between near-degenerate mass eigenstates, $\delta m^2$. 
Similarly, in Majoron models~\cite{Nussinov:1987pc,Bertolini:1987kz,Fuller:1988ega}, lepton number is spontaneously broken, generating neutrino masses, as well as a Goldstone boson (the Majoron). 
In these scenarios, neutrinos can decay on timescales dependent on the Majoron model parameters, introducing a new length scale: the neutrino lifetime~$\tau$.
Phenomenologically, the QD and Majoron models depend similarly on the ratio of the neutrino propagation distance $L$ to the energy $E$, and could manifest on energy and length scales much larger than can be probed by terrestrial neutrino sources. 

\begin{figure}
  \includegraphics[width=\textwidth]{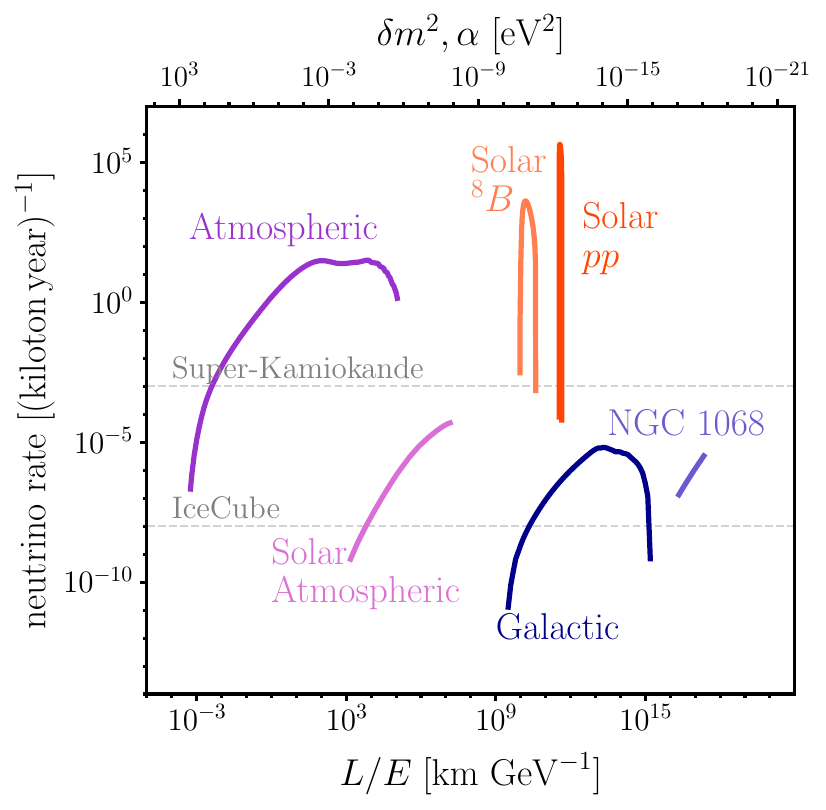}
  \caption{
  \textbf{\emph{Expected rates of neutrinos from natural sources as a function of L/E.}} 
  This figure depicts the rate of neutrinos from various natural sources, at ideal Earth-based detectors.
  We plot the rate as a function of the baseline to energy ratio $L/E$, in units of neutrinos per kiloton-year.
  For reference, the exposure of Super-Kamiokande and IceCube are indicated by grey dotted lines.
  To calculate the rate, we multiply the source flux by the relevant cross-section; for solar neutrinos, we use the neutrino-electron scattering cross-section, whereas for all other sources we use neutrino-nucleon cross-section.
  All sources except the solar atmospherics have been detected in the ranges shown. 
  }
  \label{fig:L-o-E}
\end{figure}

The observation of the Milky Way galaxy in neutrinos by the IceCube Neutrino Observatory~\cite{IceCube:2023ame} has opened up new opportunities in neutrino astronomy and particle physics. 
As shown in~\Cref{fig:L-o-E}, Galactic neutrinos occupy $L/E$ parameter space that is yet unexplored. 
Galactic neutrinos thus present an exciting opportunity to probe new neutrino physics that alters propagation on $L/E$ scales around $10^{13}\,\mathrm{eV}^{-2}$. 

In this article, we explore the sensitivity of Galactic neutrino measurements to quasi-Dirac neutrino and neutrino decay scenarios.
We forecast the sensitivity of gigaton-scale neutrino telescopes IceCube and KM3NeT~\cite{KM3Net:2016zxf} to these scenarios. 
We demonstrate that future analyses of Galactic neutrinos can constrain QD and neutrino decay models in parameter space previously unexplored by laboratory experiments.

The rest of this article is organized as follows.
First, in~\cref{sec:models}, we review the models considered.
In~\cref{sec:galactic-emission}, we next describe the production, propagation, and detection of Galactic neutrinos in our beyond the Standard Model (BSM) scenarios.  
In~\cref{sec:analysis}, we discuss details of our analysis implementation, including our treatment of detector responses, backgrounds, and our statistical methodology. 
Finally, in~\cref{sec:results}, we present our main results, which forecast the sensitivity of neutrino telescopes to QD and neutrino decay scenarios, and in~\cref{sec:conclusion}, we conclude.

\section{Models\label{sec:models}}

\begin{figure*}[ht]
    \includegraphics[width=\textwidth]{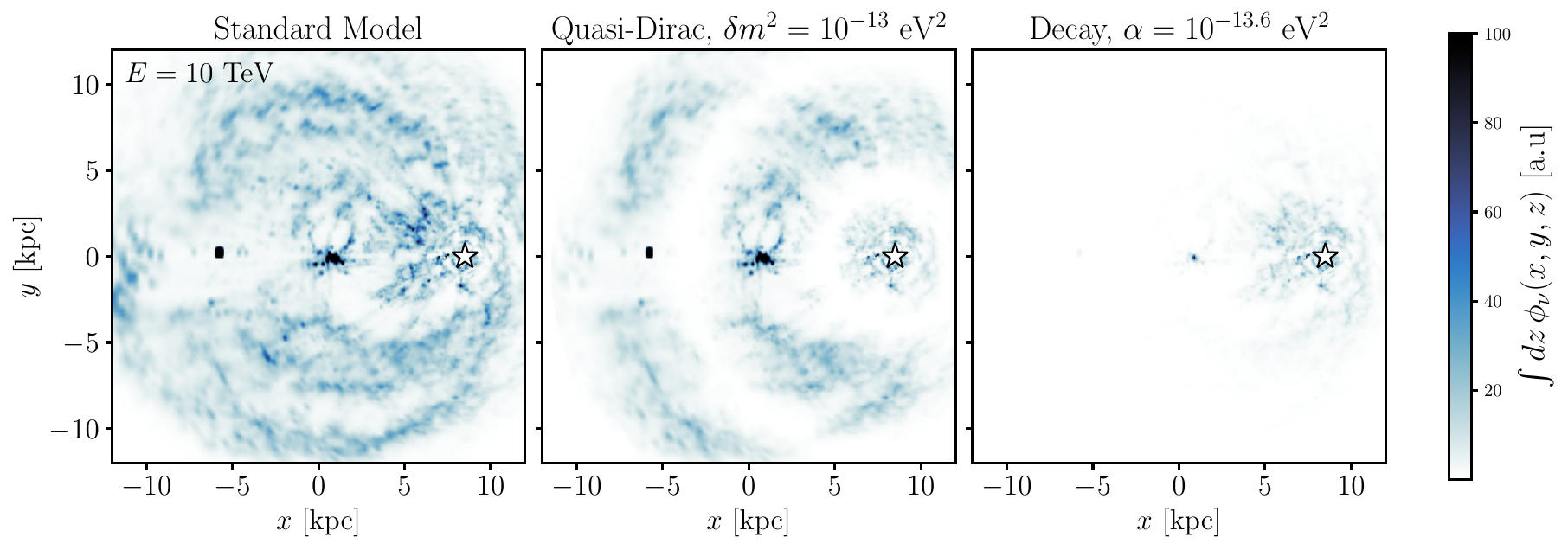}
    \caption[]{\textbf{\textit{
        Spatial distribution of Galactic neutrinos under SM and BSM scenarios.
    }} 
    Shown are maps of $z$-integrated Galactic neutrino emission at $10\,\mathrm{TeV}$ weighted by survival probability at Earth under a given BSM scenario (see~\Cref{eq:QDinos,eq:invisibledecays}). The emission model assumes a gas distribution as calculated in Ref.~\cite{Soding:2025scd}. 
    \textit{Left:} Neutrino emission under a SM scenario. 
    \textit{Middle:} Neutrino emission under a quasi-Dirac scenario with $\delta m^2 = 10^{-13}\,\mathrm{eV^2}$. 
    \textit{Right:} Neutrino emission, after decay with $\alpha = 10^{-13.6}\,\mathrm{eV^2}$. 
    }
  \label{fig:galactic-neutrino-prod}
\end{figure*}

Galactic neutrinos must propagate very long distances to reach Earth-based detectors, and can therefore be used to probe new regions in $L/E$, see~\Cref{fig:L-o-E}.
In this article, we consider two scenarios that produce observable effects for neutrinos that propagate Galactic baselines: quasi-Dirac neutrinos and neutrino decay.

If right-handed neutrinos participate in the neutrino mass mechanism---e.g. if neutrino masses are generated through a coupling with the Higgs field, as occurs for the other fermions---the mass spectrum of neutrinos will depend also on the right-handed neutrino mass term, which is a free parameter. 
In the limit in which the right-handed mass term is much smaller than the Dirac mass term, neutrinos are ``quasi-Dirac." 
This scenario appears when new physics breaks lepton number at high scales, producing a small Majorana mass, see e.g.~\cite{Silagadze:1995tr,Joshipura:2013yba,Gu:2006dc,Ma:2014qra,Valle:2016kyz,CentellesChulia:2018bkz,Lee:1956qn,Foot:1991py,Berezhiani:1995yi}.
In the quasi-Dirac scenario, there are six neutrino mass eigenstates, grouped in three quasi-degenerate pairs, each with a hyperfine splitting $\delta m^2_i$.
For the $L/E$ of galactic neutrinos, the known mass differences $\Delta m^2_{32}, \Delta m^2_{21}$ can be averaged out, such that the probability of a neutrino $\nu_\alpha$ of flavor $\alpha$ oscillating into a neutrino $\nu_\beta$ of flavor $\beta$ is given by
\begin{equation}
\label{eq:QDinos}
    P_{\alpha\to\beta}^\mathrm{QD} = \sum_{i=1}^3 |U_{\alpha i}|^2 |U_{\beta i}|^2 \cos^2 \left( \frac{\delta m_i^2 L}{4 E}\right),
\end{equation}
where the sum runs over the three mass eigenstate pairs and $U$ is the PMNS matrix.

Solar neutrinos constrain QD mass splittings across many orders of magnitude: $\delta m_1^2$ is constrained down to $10^{-12}\,\mathrm{eV^2}$ and $\delta m_2^2$ down to $\sim 10^{-11}\,\mathrm{eV^2}$ at $3\sigma$ confidence level~\cite{deGouvea:2009fp, Ansarifard:2022kvy}. 
Measurements of the $pp$ solar neutrino flux at the JUNO experiment can reach $\delta m^2 \sim 10^{-13}\,\mathrm{eV^2}$~\cite{Franklin:2023diy}. 
Neutrinos from SN1987A have also been used to search for QD neutrinos, producing constraints on $\delta m^2 \in [2, 4]\times 10^{-20}\,\textrm{eV}^2$ at the $2\sigma$ confidence level~\cite{Martinez-Soler:2021unz}.
Finally, the spectrum of high-energy all-sky astrophysical neutrinos has been recently used to constrain $\delta m^2 \in [5, 8]\times 10^{-19}\,\textrm{eV}^2$ at the $3\sigma$ level, assuming the redshift distribution of astrophysical sources follows the star formation rate density~\cite{Carloni:2025dhv}.
Other proposed searches for QD neutrinos include analyses of the diffuse supernova neutrino background (DSNB)~\cite{DeGouvea:2020ang}, cosmogenic neutrinos~\cite{Leal:2025eou}, and high-energy neutrinos from astrophysical point sources~\cite{Beacom:2003eu, Carloni:2022cqz}.

Many models extending the SM predict neutrino decay~\cite{Bahcall:1972my,Shrock:1974nd,Petcov:1976ff,Marciano:1977wx,Zatsepin:1978iy,Chikashige:1980qk,Gelmini:1980re,Pal:1981rm,Schechter:1981cv,Shrock:1982sc,Gelmini:1983ea,Bahcall:1986gq,Nussinov:1987pc,Frieman:1987as,Kim:1990km}. The decay products could be lighter neutrino mass states or new invisible particles. 
Here, we take a model agnostic approach, for which the only relevant parameters of the decay are the mass-to-lifetime ratios $\alpha_i \equiv m_i/\tau_i$ of the neutrino mass states $m_i$.

In this work, we consider two main classes of decay scenarios. 
In the first, termed \textit{invisible decays}, neutrinos decay entirely to undetectable daughters.
In this scenario, a Galactic neutrino of flavor $\alpha$ and energy $E$ has lifetime $\tau_i E / m_i = E/\alpha_i$, and has probability of reaching Earth in flavor $\beta$ given by:
\begin{equation}
\label{eq:invisibledecays}
    P_{\alpha \to \beta}^\mathrm{decay} = \sum_{i=1}^3 |U_{\alpha i}|^2 |U_{\beta i}|^2\exp\left\{ - \frac{\alpha_iL}{E_\nu}  \right\},
\end{equation}
where in this equation, the sum is taken over the three mass eigenstates. 
It is also possible for neutrinos to undergo two-body decay into a lighter neutrino mass state and another exotic decay product. 
These processes are called \textit{visible decays}, and yield increased fluxes of light neutrinos.

The strongest constraints on neutrino lifetimes come from astrophysical sources and cosmological observations. Usually, these constraints are model dependent and/or only apply to certain neutrino mass states. 
Solar neutrinos can bound $\alpha \gtrsim 10^{-13}~\mathrm{eV^2}$ for invisible $\nu_1$ and $\nu_2$ decays~\cite{Berryman:2014qha} and certain visible decay scenarios~\cite{Picoreti:2021yct}. 
The diffuse astrophysical neutrino flux detected by IceCube can set bounds up to $\alpha \gtrsim 10^{-18}~\mathrm{eV^2}$ on invisible neutrino decays if both $\nu_2$ and $\nu_3$ decay~\cite{Valera:2024buc}, and SN1987A has been used to set model-dependent bounds on decay parameters up to $\alpha \sim 10^{-21}~\mathrm{eV}^2$~\cite{Ivanez-Ballesteros:2023lqa, Martinez-Mirave:2024hfd}. 
Cosmology can also act as an indirect and complementary probe of neutrino decay~\cite{Escudero:2019gfk, Barenboim:2020vrr}, setting bounds down to $\alpha \lesssim 10^{-26}\,\mathrm{eV^2}$, although more recent work has weakened the bounds to $\alpha \lesssim 10^{-22}\,\mathrm{eV^2}$~\cite{Chen:2022idm} and the bounds are dependent on the absolute neutrino mass scale. 
Further studies have explored the possibility of using detections of high-energy astrophysical neutrino point sources~\cite{Valera:2024buc} and the DSNB~\cite{Martinez-Mirave:2024hfd, MacDonald:2024vtw, Ivanez-Ballesteros:2022szu} to probe both visible and invisible neutrino decay scenarios.

Throughout this work, we will denote individual BSM parameters––i.e. QD mass-squared differences $\delta m_i^2$ or decay parameters $\alpha_i$––with mass state indices $i$. 
BSM parameters without indices, $\delta m^2$ or $\alpha$, refer to a value common across all three mass states.

\section{Galactic neutrino production, propagation, and detection\label{sec:galactic-emission}}

In this section, we discuss our implementation of Galactic neutrino emission, propagation, and relevant detector resolutions.

\textbf{\textit{Production:}} High-energy neutrinos are produced alongside gamma rays in the collisions of high-energy cosmic rays (CRs) with the gas in our galaxy, generating a ``diffuse'' flux. 
Additionally, observations of PeVatrons in the Galaxy~\cite{HESS:2016pst,LHAASO:2021gok} suggest the existence of localized high energy neutrino sources.
Thus far, IceCube has found evidence at greater than $4\sigma$ for neutrino emission spatially correlated with multiple models of the diffuse emission~\cite{IceCube:2023ame}, although a fraction of this observed flux could come from unresolved Galactic neutrino sources~\cite{Schwefer:2022zly, Fang:2023azx, DeLaTorreLuque:2025zsv, Marinos:2025ddy}.

In this work, we focus on the diffuse component of the Galactic neutrino flux.
We use the \texttt{TANDEM} model suite~\cite{Carloni:2025upc} for this diffuse flux, since it is designed to describe the detailed three dimensional distribution of neutrino production.
These models are generated from the product of a variety of gas maps with CR distributions produced using \texttt{CRPropa}~\cite{CRPropa:2022ovg}.
\texttt{TANDEM} separately computes the neutrino spectrum by integrating the CR fluxes from the \texttt{Global Spline Fit}~\cite{Dembinski:2017zsh} against differential neutrino production cross-sections from \texttt{AAfrag}~\cite{Koldobskiy:2021nld, Kachelriess:2019ifk}.
Additional unresolved sources could contribute to the total Galactic flux with a different spectrum, but a likely similar spatial distribution; we leave the detailed treatment of such localized sources to future work, but account for this potential component by leaving the flux normalization unconstrained.

\textbf{\textit{Propagation:}} To incorporate BSM propagation effects, we integrate the four-dimensional neutrino emission predictions from \texttt{TANDEM} against \Cref{eq:QDinos,eq:invisibledecays} for each line of sight.
For example, the neutrino flux in a QD scenario at a given Galactic coordinate $(\ell, b)$ is given as 
\begin{equation}
    \begin{aligned}
        \phi_\alpha (E_\nu, &\ell, b) = \sum_{\beta, i} |U_{\alpha i}|^2|U_{\beta i}|^2 \\ &\times \int_0^\infty dr~F_\beta(E_\nu, r, \ell, b)\cos^2\left(\frac{\delta m_i^2 r}{4E_\nu}\right),
    \end{aligned}
\end{equation}
where $F_\beta$ is the spectral emissivity of neutrinos of flavor $\beta$ per unit line-of-sight distance obtained from~\texttt{TANDEM}, \[ F_\beta = \frac{dN^\nu_\beta}{dE_\nu\,d\Omega\,dt\,dA\,dr}. \]
The calculation for invisible decay models is similar, while visible decay scenarios require a more involved expression, described in \cref{sec:appvisdecay}. 

We show the effect of these BSM scenarios on the neutrino spatial distribution, compared to the SM prediction of a reference \texttt{TANDEM} model, in \Cref{fig:galactic-neutrino-prod}. 
As can be seen in the figure, the geometry of our Galaxy implies that the typical propagation distance $L$ depends strongly on the line-of-sight direction. 
Neutrinos coming from the Galactic center typically travel distances on the order of 10\,kpc, whereas neutrinos from the opposite direction travel only a few\,kpc. 
When integrated over a distribution in propagation distance $L$, the $L/E$ dependent \textit{survival} probabilities of~\Cref{eq:QDinos,eq:invisibledecays} produce observable modifications to the energy spectrum.
Due to the variation in typical propagation distances, these spectral distortions are typically direction-dependent.

\begin{figure}
    \includegraphics[width=\textwidth]{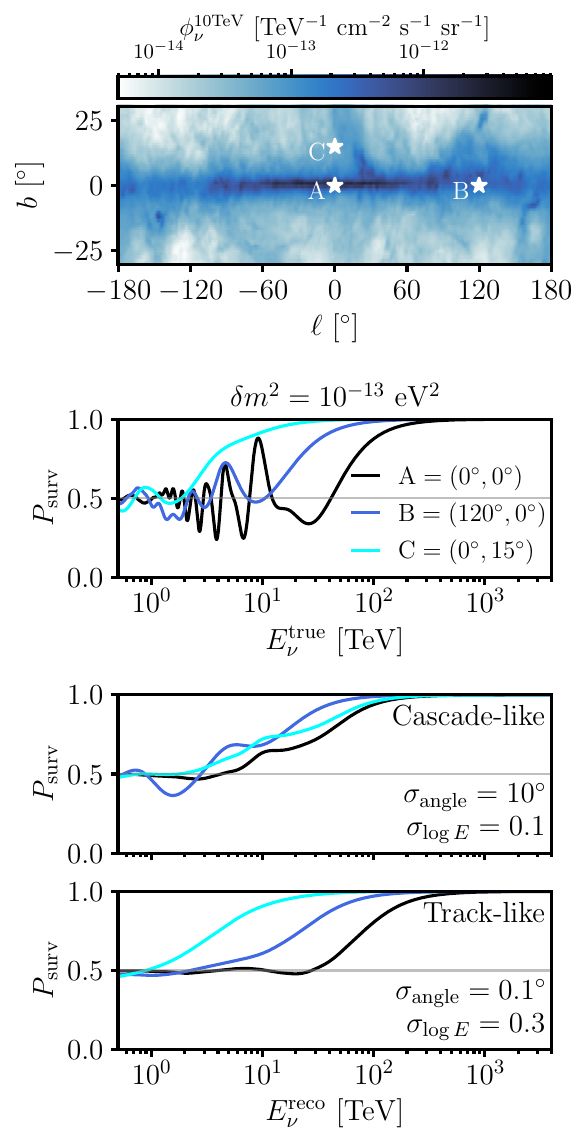}
    \caption[]{\textbf{\textit{Survival probability of neutrino emission along different lines of sight in a quasi-Dirac scenario.}} 
\textit{Top:} Galactic neutrino flux as a function of Galactic latitude $\ell$ and longitude $b$, with three representative sky locations marked: A at $(0^\circ, 0^\circ)$, B at $(60^\circ, 0^\circ)$, and C at $(0^\circ, 15^\circ)$. 
    \textit{Middle:} ``Survival" probability, i.e. the probability that neutrinos along a given line of sight remain in active states, as a function of true neutrino energy.
    \textit{Bottom:} ``Survival" probability, accounting for energy and angular resolutions characteristic of cascade (above) and track (below) events.
    }
  \label{fig:oscsmears_QD_soding}
\end{figure}

\begin{figure}
    \includegraphics[width=0.93\textwidth]{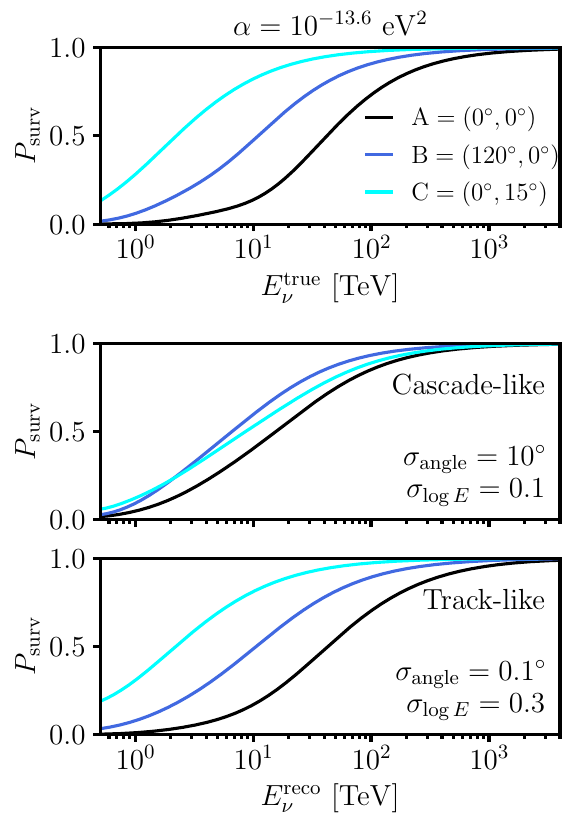}
    \caption[]{\textbf{\textit{Survival probability of neutrino emission along different lines of sight in a decay scenario.}} Same as Fig. \ref{fig:oscsmears_QD_soding} (middle, bottom), but for an invisible decay scenario with $\alpha = 10^{-13.6}~\mathrm{eV^2}$.
    }
  \label{fig:oscsmears_decay_soding}
\end{figure}

\textbf{\textit{Detection:}} In this work, we focus on complementary measurements of Galactic neutrinos at IceCube and KM3NeT/ARCA.\footnote{When only KM3NeT is written, we refer to KM3NeT/ARCA.}
Both telescopes detect neutrinos by looking for the Cherenkov light produced by the charged products of neutrino interactions.
Events in these detectors are typically classified into two morphologies, based on the pattern of light deposition. 
\textit{Tracks} are long linear depositions of light, predominantly produced by charged-current $\nu_\mu$ interactions.
\textit{Cascades} are compact spherical light depositions produced by charged-current $\nu_e$ and $\nu_\tau$ interactions, as well as neutral-current interactions of all flavors.

Thus far, IceCube has only reported evidence of neutrinos from the Galaxy using cascade-based analyses.
Searches for Galactic neutrinos using track events in IceCube are typically difficult. Because the Galaxy is located in the Southern sky, and IceCube at the South Pole, the majority of Galactic neutrinos arrive at IceCube from above the horizon, where atmospheric muons produce an enormous background for track-based searches.
However, since KM3NeT is located in the Mediterranean, Galactic neutrinos propagate through the Earth to arrive at the detector; muons cannot penetrate the Earth, and thus the primary background for track-based searches is atmospheric muon neutrinos, at a lower rate.

Track and cascade events also have different, and complementary, reconstruction properties. 
Due to the extension of track events through the detector volume, track events will naturally achieve more precise angular resolution, typically below 1$^\circ$ above a few~TeV. 
On the other hand, because the muon trajectory typically extends beyond the instrumented detector volume, and the muon loses energy stochastically, only a fraction of its energy is observable, resulting in poor energy reconstruction.
Conversely, in cascade events, the products of the neutrino interaction deposit most of their energy within the instrumented detector volume, leading to better energy resolution; but due to their compact morphology and shorter lever arm, they have poor angular resolution, on the order of 10$^\circ$ for ice-based Cherenkov detectors and 2$^\circ$ for water-based~\cite{IceCube:2020wum,KM3NeT:2024paj}.

We illustrate the differences in the observable QD signatures for tracks and cascades, along different lines-of-sight, in~\Cref{fig:oscsmears_QD_soding}, for a reference \texttt{TANDEM} model and mass-squared splitting $\delta m^2 = 10^{-13}\,\mathrm{eV}^2$.
In the top panel, we show the overall spatial distribution of the Galactic neutrino flux predicted by the model, and mark three example directions: the first pointing towards the Galactic center (A), the second remaining on the Galactic plane but away from the center (B), and the third pointing off the plane at high-latitudes (C). 
In the second panel, we show the line-of-sight integrated neutrino survival probability, i.e. the probability that neutrinos remain detectable after propagation, as a function of true energy $P_\textrm{surv}(E)$ in each of these three directions.
As expected, at high energies there is no disappearance effect, while at low energies the survival probability approaches a half, due to the integration averaging the effects of many high-frequency oscillations.
Additionally, because neutrinos from the Galactic center travel on average much longer distances than neutrinos from larger longitudes, the first oscillation minimum occurs at higher energies: around 30\,TeV in case A, versus 10\,TeV in case B. 

\begin{table}[tp]
    \centering
    \renewcommand{\arraystretch}{1.15}
    \begin{tabularx}{\textwidth}{@{} X S[table-format=+1.2] S[table-format=+1.2] @{}}
        \toprule
        & \multicolumn{2}{c}{\textbf{Relative normalization shift (\%)}} \\
        \cmidrule(lr){2-3}
        \textbf{BSM scenario} & {\textbf{Cascades}} & {\textbf{Tracks}} \\
        \midrule
        $\nu_3 \to \mathtt{invis}$ & -28 & -47 \\
        $\nu_2 \to \mathtt{invis}$ & -30 & -41 \\
        $\nu_1 \to \mathtt{invis}$ & -41 & -13 \\
        \addlinespace
        $\nu_3 \to \nu_1$ & +13 & -34 \\
        $\nu_3 \to \nu_2$ & +2 & -6 \\
        $\nu_2 \to \nu_1$ & +11 & -28 \\
        \addlinespace
        quasi-Dirac & -50 & -50 \\
        \bottomrule
    \end{tabularx}    
    \caption{\textbf{\textit{Effect on the normalization of the Galactic neutrino flux to BSM scenarios in the $\delta m^2, \alpha \gg E/L$ limit.}} The BSM scenarios considered are invisible decays, two-body visible decays, and quasi-Dirac oscillations (equal $\delta m^2$). 
    }
    \label{tab:normeffects}
\end{table}

Finally, in the bottom two panels of~\Cref{fig:oscsmears_QD_soding}, we plot the same survival probabilities after convolution with detector responses typical of cascade or track event morphologies. 
For the cascade-like scenario, we use a broad angular resolution of $10^\circ$, but a narrow energy resolution of 0.1 in log-space. 
This energy resolution produces survival probabilities that retain some small-scale features, but the poor angular resolution smears directions pointing towards the plane with those looking at higher-latitudes, reducing the large-scale differences between the three curves.
In contrast, in the track-like scenario we use an angular resolution  of $0.1^\circ$ but poorer energy resolution, 0.3 in log-space. 
The resulting curves have no small-scale oscillation features, but remain much more distinct.

In~\Cref{fig:oscsmears_decay_soding} we similarly show the neutrino survival probability for each of these lines-of-sight in an invisible decay scenario with decay parameter $\alpha = 10^{-13.6}\,\mathrm{eV}^2$.
Because the decay channel lifetime~$\tau$ is boosted in the laboratory frame by the Lorentz factor $E/m$, the survival probability approaches zero at low energies, where the lab-frame lifetime is short. 
High-energy neutrinos from the Galactic center, which travel on average longer distances, are more likely to decay before reaching Earth than neutrinos from other directions.
As in the QD case, the poor angular resolution of the cascade-like scenario smears out these directionally-dependent effects.

Lastly, we note that in the limits in which $\delta m^2, \alpha \gtrsim 1\,\textrm{PeV}/1\,\textrm{kpc} \simeq 10^{-11}\,\textrm{eV}^2$, both the QD and decay scenarios produce similar effects across all energies, resulting in normalization shifts instead of shape effects. 
If the normalization of the Galactic neutrino flux can be constrained from model predictions, then the entirety of these regions of parameter space can be probed.
This possibility is particularly interesting for decay scenarios, which produce much more dramatic normalization changes.
In \Cref{tab:normeffects}, we list the relative normalization shifts in the total number of cascade or track events for a variety of invisible and visible decay scenarios, as well as for a QD scenario with a single mass-squared splitting, in this \textit{large parameter regime}. 
For this table, we assume that the neutrino mass states are normally ordered.
Scenarios such as the decay $\nu_3 \to \nu_1$ produce very different normalization shifts for cascade-based or track-based measurements of Galactic neutrinos; a joint analysis using these two channels is therefore much more sensitive to this scenario than either measurement individually.
QD models with distinct squared-mass splittings $\delta m^2_i$ produce similar effects, except smaller in magnitude.

\section{Analysis~\label{sec:analysis}}

\begin{figure*}[t!]
        \includegraphics[width=\textwidth]{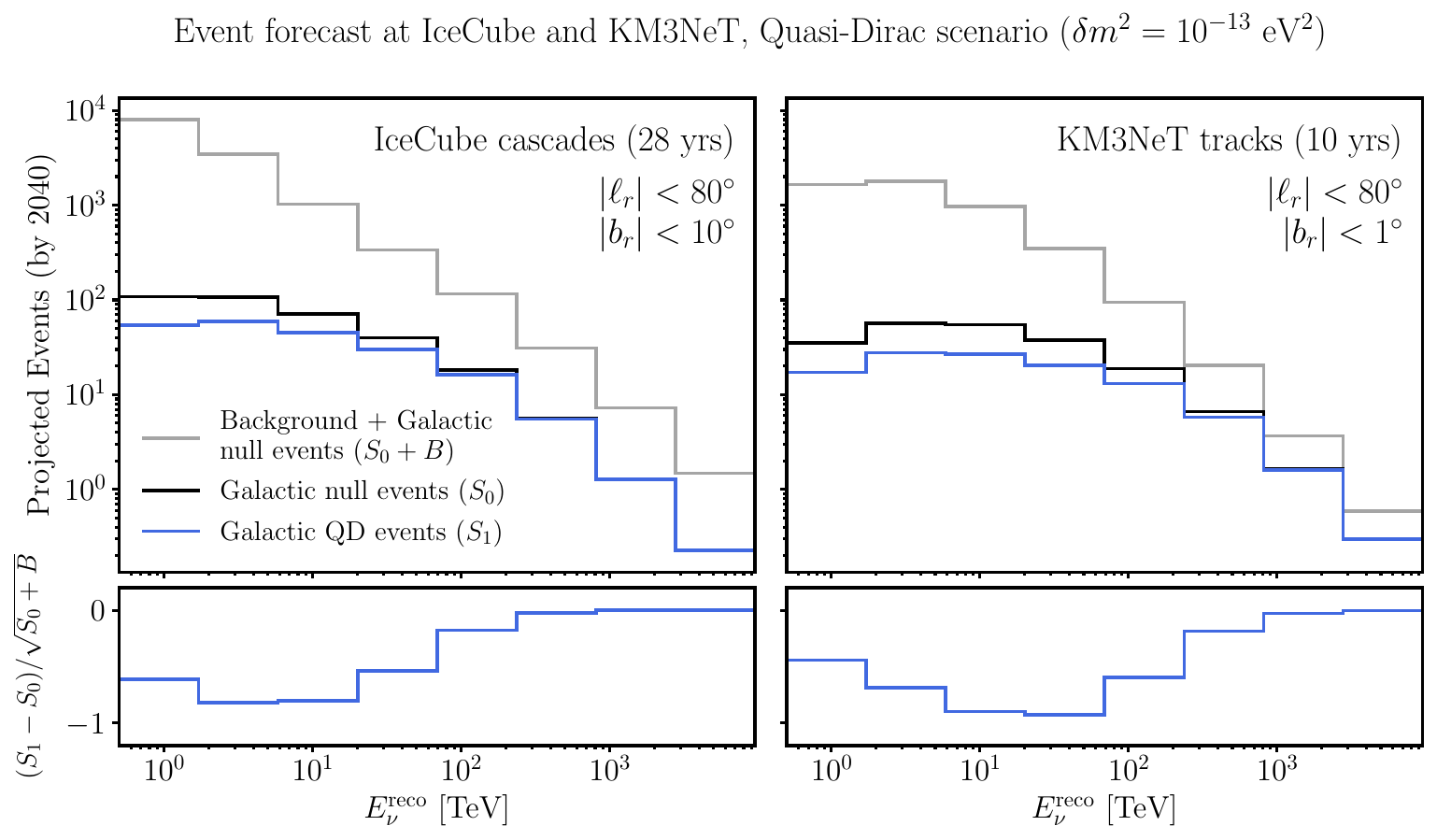}
        \caption[]{\textbf{\textit{Predicted distributions of events from the Galactic center region in IceCube and KM3NeT under SM and QD scenarios.}} 
        On the left, we show all IceCube cascade events within a wide Galactic window, ($|\ell| < 80^\circ,~|b| < 10^\circ$), whereas on the right we show KM3NeT track events in a narrower window ($|\ell| < 80^\circ,~|b| < 1^\circ$).    
        \emph{Top rows:} Distributions of Galactic neutrinos in energy in the SM ($S_0$) and QD ($S_1$) scenarios, as well as of background atmospheric and diffuse astrophysical neutrinos ($B$). 
        \emph{Bottom rows:} Ratio of the QD disappearance effect ($S_1 - S_0$) to the statistical uncertainty on the total event count, in each energy bin.
        }
      \label{fig:binned_event_rates_inner_window}
    \end{figure*}

In our study, we consider a single QD scenario, with a unique squared-mass splitting $\delta m^2_i = \delta m^2\,\forall i$, and two neutrino decays, both involving an unstable $\nu_3$ mass state: the first an \textit{invisible decay} to unspecified undetectable daughters, and the second a visible decay to $\nu_1$.
For our visible decay scenario, we consider only the quasi-degenerate limit; see~\Cref{sec:appvisdecay} for more discussion.
We assume that the neutrino mass states are normally ordered, i.e. that $\nu_3$ is the heaviest mass state and $\nu_1$ the lightest, and we use the PMNS matrix entries from Ref.~\cite{Esteban:2024eli}.
We also assume a $(\nu_e:\nu_\mu:\nu_\tau) = (1:2:0)$ flavor ratio at production, since we consider only the diffuse production of galactic neutrinos from the decays of pions produced in CR-gas interactions.
We fix the normalization of our Galactic neutrino flux such that it predicts 650 cascade events over 10 years at IceCube, which is consistent with the IceCube results~\cite{IceCube:2023ame}.

We project the sensitivity of a future (2040) analysis, assuming 28 years of IceCube measurements based on cascade events and 10 years of full KM3NeT/ARCA measurements using track events.
In principle, KM3NeT will also be sensitive to Galactic neutrinos in analyses based on both cascade events. 
We do not examine the cascade channel at this time, since one of our goals is to emphasize the complementarity of track and cascade measurements, but doing so should increase sensitivity.
To calculate event rates at the two detectors, we multiply the neutrino fluxes at Earth by energy- and declination-dependent effective areas, and then convolve with the detector resolutions. 
We model the IceCube detector response using the effective areas and angular and energy resolutions from Ref.~\cite{IceCube:2023ame}. We model the KM3NeT response using the effective areas from Ref.~\cite{KM3NeT:2024paj}, an angular resolution of $0.1^\circ$, and an energy resolution of 0.3 in log-space~\cite{Margiotta:2022kid}. 
To estimate the effect of backgrounds on the sensitivity, we model atmospheric neutrino backgrounds using the \texttt{H3a\_SIBYLL23C} model from \texttt{nuflux}. 
For downgoing directions, we implement the muon self-veto effect using the tabulated values in Ref.~\cite{Arguelles:2018awr}. 
We model the diffuse astrophysical neutrino flux as a broken power-law, using the best-fit result from Ref.~\cite{IceCube:2025tgp}.

The distribution of events from the Galactic center direction over energy is shown in~\Cref{fig:binned_event_rates_inner_window}, along with the distribution of Galactic neutrino events in a QD scenario with $\delta m^2 = 10^{-13}\,\textrm{eV}^2$. 
For the IceCube cascade channel, we plot all events in a large angular window centered on the Galaxy, with width of $160^\circ$ in longitude and $20^\circ$ in latitude.
For the KM3NeT track channel, which has much better angular resolution, we plot the events in a narrower angular window, with equal width in longitude but only $2^\circ$ height in latitude.
The total event distribution, which is the sum of atmospheric and Galactic neutrino contributions, is plotted in gray.
While the total rate of atmospheric backgrounds is larger for the track channel, since the atmospheric $\nu_\mu$ flux is an order of magnitude greater than the $\nu_e$ flux at TeV energies, the better angular resolution reduces the solid angle over which Galactic neutrino events are spread, reducing the effective background rate.
The Galactic neutrino distribution is plotted in black in the baseline SM case, and in blue for the QD case.
In the lower axes, we plot the ratio of the QD disappearance effect, given by the difference in the two Galactic neutrino predictions, to the statistical uncertainty on the total event rate.
In both the IceCube cascades and KM3NeT track scenarios, we find that the magnitude of the QD signature is at best on the order of magnitude of the total statistical uncertainties, indicating that extant backgrounds in the Galactic neutrino detection significantly limit sensitivity.

To compute the sensitivity of IceCube and KM3NeT/ARCA measurements of the Galactic neutrino emission to our three BSM models, we perform a binned likelihood ratio test, assuming Poisson statistics. 
We define the following test statistic:
\begin{equation}
    \begin{aligned}
        -&2\Delta\mathrm{LLH} = \min_{\xi,\eta}\; 2\sum_i \Biggl\{
        (1+\xi)N_i^G  - \mu^G_i + \eta \mu_i^\mathrm{astro} \\
        &+ (\mu^G_i \left.+ \mu^\mathrm{atm}_{i} +\mu_i^\mathrm{astro}\right) \\&\quad\times\log\left(\frac{\mu^G_i + \mu^\mathrm{atm}_{i}+\mu^\mathrm{astro}_{i}}{(1+\xi)N_i^G + \mu^\mathrm{atm}_{i}+(1+\eta)\mu_i^\mathrm{astro}}\right)\Biggr\} \\ &+ \left(\frac{\eta}{\sigma_\mathrm{astro}} \right)^2. \\
    \end{aligned}
\end{equation}
In this expression, $\mu^G_i$ is our nominal SM signal event count, $\mu^\mathrm{atm}_i$ is our nominal atmospheric background event count, $\mu^\mathrm{astro}_i$ is our nominal diffuse astrophysical event count, and $N_i^G$ is the number of BSM signal events predicted by the model; the summation index $i$ runs over all energy and angular bins. 
$\xi$ is a nuisance parameter that accounts for the systematic uncertainty on the overall GP neutrino flux normalization, and $\eta$ is a pull parameter that accounts for the normalization uncertainty $\sigma_\mathrm{astro}$ on the diffuse astrophysical neutrino flux \cite{IceCube:2025tgp}. 

Since there are still substantial uncertainties on the total magnitude of the Galactic neutrino flux, we leave $\xi$ as a free parameter in our fit and profile over it in our analysis, and we also profile over $\eta$ in the fit. 
We use an energy binning of 0.16 in log-space for both IceCube and KM3NeT, and we use square angular bins in Galactic coordinates with widths $7^\circ$ for IceCube and $1^\circ$ for KM3NeT.

\section{Results~\label{sec:results}}

\begin{figure*}
    \includegraphics[width=\textwidth]{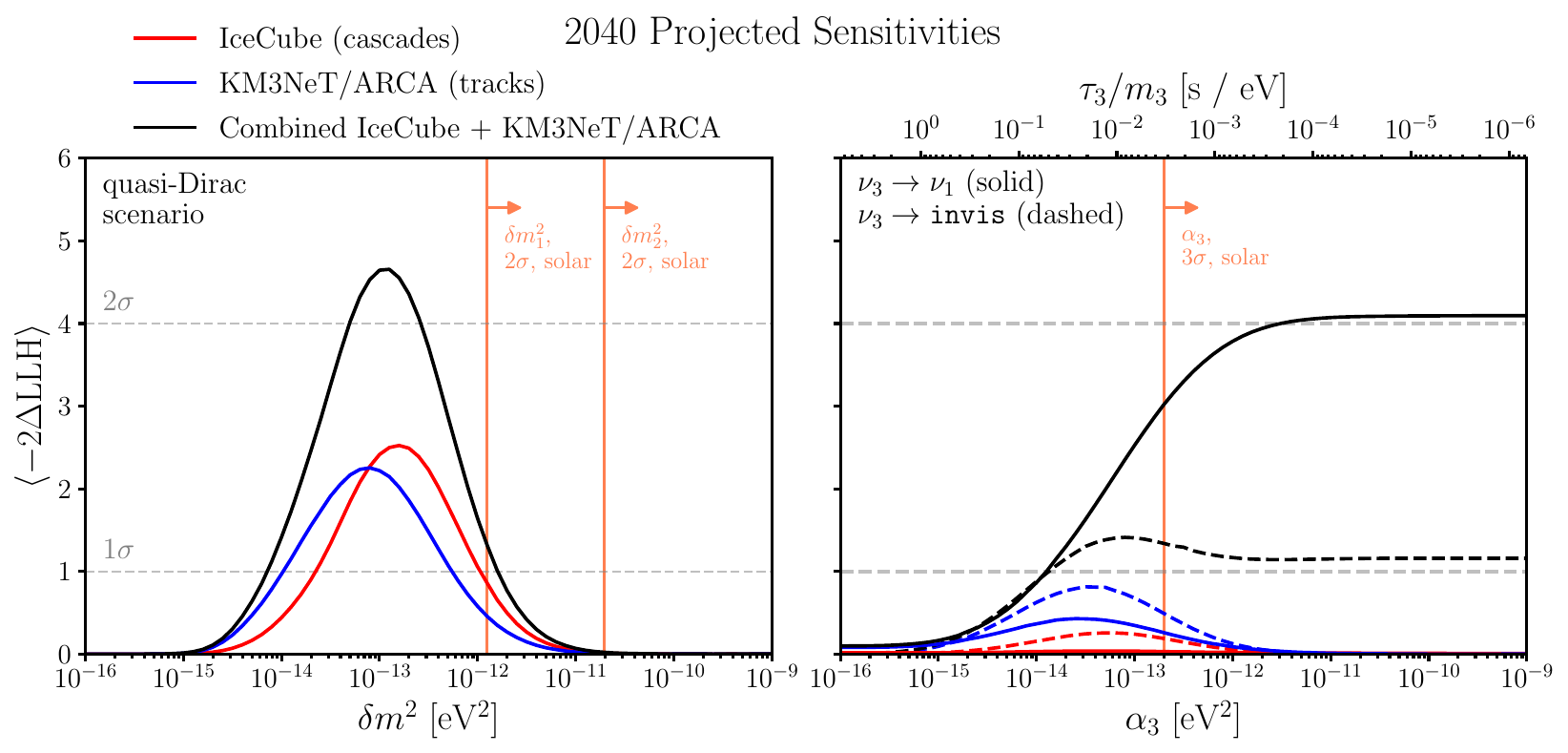}
    \caption{
        \textbf{\emph{Sensitivity to QD and neutrino decay models.}}
        \textit{Left:} The median test statistic for IceCube cascades (red) and KM3NeT/ARCA tracks (blue) analyses as a function of $\delta m^2$. 
        \textit{Right:} Same as left, but for the decay scenarios $\nu_3 \to \nu_1$ (solid) and $\nu_3 \to \mathtt{invis}$ (dashed), plotted as a function of the decay parameter $\alpha_3$ (bottom axis) and the neutrino lifetime-over-mass ratio $\tau_3 / m_3$ (top axis).
        In both figures the solid black line corresponds to the combined analysis. 
        Sensitivities are projected for 2040, assuming 28 years of IceCube and 10 years of the full KM3NeT/ARCA detector. 
        Relevant bounds from solar neutrinos are shown in vertical orange lines; the quasi-Dirac bounds are taken from Ref.~\cite{Ansarifard:2022kvy} and the decay bound is taken from Ref.~\cite{Picoreti:2021yct}. 
        The $\alpha_3$ invisible decay parameter space is excluded by astrophysical neutrino flavor ratio measurements under certain source model assumptions by Ref.~\cite{Valera:2024buc}.
    }
    \label{fig:sensitivities}
\end{figure*}

In \Cref{fig:sensitivities}, we present the Asimov exclusion sensitivities to quasi-Dirac and decay scenarios in a forecast 2040 analysis. 
For our main result, we use a \texttt{TANDEM} model based on the gas model in Ref.~\cite{Soding:2025scd}, the \texttt{UF23.base} Galactic magnetic field model in Ref.~\cite{Unger:2024cmf}, and a CR source distribution that follows the Galactic supernova remnant population, implemented in Ref.~\cite{CRPropa:2022ovg}. 
We explore the effect of changing the neutrino emission model in Appendix~\ref{sec:appdiffgasmodels}.
We consider both the individual sensitivities and the combined sensitivities of both experiments. 

For the QD model, we find that a combined IceCube-KM3NeT analysis of the Galactic neutrino flux is sensitive to squared-mass splittings $\delta m^2 \in [2.5\times 10^{-14}, 5 \times 10^{-13}]\,\textrm{eV}^2$ at the $90\%$ confidence level.
Due to its larger exposure, IceCube alone is more sensitive than KM3NeT.
The offset in the peak sensitivity of the IceCube and KM3NeT results is due to differences in the detector energy threshold, which is lower for KM3NeT.

In both the decay scenarios considered, we find that neither experiment is individually sensitive to $\nu_3$ decay effects, but a joint analysis is considerably more powerful than the sum of sensitivities obtained by individual experiments. 
This is because the flavor-dependence of the decay effect produces different signatures in the cascade and track samples, including in the large-parameter limit, where the decay signature shifts from a spectral to a normalization-only effect.
Thus, while the single experiment sensitivities are substantially reduced by profiling over the unknown Galactic flux normalization, the joint analysis sensitivity is not. 
This effect is more dramatic in the decay scenario $\nu_3 \to \nu_1$ than the invisible, since the asymmetry in cascade and track signatures is larger in this case (see~\Cref{tab:normeffects}). 

Our result finds that a joint IceCube-KM3NeT analysis of the Galactic neutrino flux is sensitive to $\nu_3 \to\ \nu_1$ decays with decay parameter $\alpha_3 > 1.6 \times 10^{-13}\,\mathrm{eV}^2$ at the $90\%$ confidence level. 
Our constrained region is similar to that excluded by Ref.~\cite{Picoreti:2021yct} at $3\sigma$ based on solar neutrino measurements for a helicity flipping $\nu_3 \to \bar{\nu}_1$ decay model.
For the invisible $\nu_3$ decay scenario, we do not reach sensitivity at the 90\% confidence level in any part of parameter space.
The accessible parameter space for future studies should extend down to $\alpha_3 \simeq 10^{-13}\,\textrm{eV}^2$, which is far below current bounds from atmospheric neutrinos~\cite{GonzalezGarcia:2008idc}.
Recent limits set using diffuse astrophysical neutrino flavor measurements constrain smaller $\alpha_3$ values, but are subject to source model assumptions~\cite{Valera:2024buc}. 

\section{Conclusions and Outlook\label{sec:conclusion}} 

In this work, we investigate for the first time the potential of Galactic neutrinos to probe new neutrino physics affecting ultra-long baseline propagation. 
We model the phenomenology of these scenarios using the \texttt{TANDEM} model, which provides information about the spatial distribution of Galactic neutrinos along any line-of-sight. 
We find that the BSM signal is directionally dependent and smeared out due to the range of baselines present in the Galaxy.

We forecast sensitivities to quasi-Dirac and neutrino decay scenarios for the IceCube and KM3NeT/ARCA experiments, considering both individual and combined analyses. 
We find that the Galactic neutrino flux is sensitive to several decades of unexplored parameter space in QD and neutrino decay scenarios. 
Our sensitivity is limited by the intrinsic smearing of $L/E$ signatures due to the spread in neutrino baselines over the Galaxy, systematic uncertainties on the Galactic flux normalization, and large backgrounds to the Galactic flux that obscure spectral features.

Future work could improve the sensitivity in multiple ways.
In our analysis, we leave the Galactic neutrino flux normalization entirely unconstrained, and profile over it.
However, a model-motivated prior on this normalization would make measurements sensitive not only to spectral, but also to normalization shift effects, substantially improving and extending the sensitivity. 
A similar normalization constraint could also be achieved by combining neutrino measurements with gamma-ray observations, leveraging their joint hadronic production mechanisms.
If Galactic neutrino point sources, which would have much better defined baselines, are discovered, they would be sensitive to new physics effects on similar propagation scales.
Alternately, improved angular resolution would reduce backgrounds in the Galactic signal region. 
Cascade-based measurements at KM3NeT, which we do not consider in this work, should have similar energy resolution to IceCube cascades, but improved angular resolution, due to the reduced scattering of light in water compared to ice; including them should thus improve the sensitivity.
Other water-based neutrino telescopes like Baikal-GVD~\cite{Baikal-GVD:2018isr}, which is currently in operation, and P-ONE~\cite{P-ONE:2020ljt}, TRIDENT~\cite{TRIDENT:2022hql}, and HUNT~\cite{Huang:2023mzt} can also contribute to a future study.

\acknowledgments
We thank Peter Denton for useful comments on the manuscript. MM acknowledges support from the Harvard College Research Program (HCRP) and funding from the Faculty of Arts and Sciences of Harvard University. 
KC is supported by the NSF Graduate Research Fellowship under Grant No. 2140743, and the Research Corporation for Science Advancement Cottrell Scholar award. 
CAA is supported by the Faculty of Arts and Sciences of Harvard University, the National Science Foundation, the Research Corporation for Science Advancement, and the David \& Lucile Packard Foundation. IMS is supported by STFC grant ST/T001011/1. RAB acknowledges support from the Agence Nationale de la Recherche (ANR), project ANR-23-CPJ1-0103-01.

\bibliography{nu-phys-galactic-bib.bib}

@article{Carloni:2022cqz,
    author = "Carloni, Kiara and Mart\'\i{}nez-Soler, Ivan and Arguelles, Carlos A. and Babu, K. S. and Dev, P. S. Bhupal",
    title = "{Probing pseudo-Dirac neutrinos with astrophysical sources at IceCube}",
    eprint = "2212.00737",
    archivePrefix = "arXiv",
    primaryClass = "astro-ph.HE",
    doi = "10.1103/PhysRevD.109.L051702",
    journal = "Phys. Rev. D",
    volume = "109",
    pages = "L051702",
    year = "2024"
}

@article{Martinez-Soler:2021unz,
    author = "Martinez-Soler, Ivan and Perez-Gonzalez, Yuber F. and Sen, Manibrata",
    title = "{Signs of pseudo-Dirac neutrinos in SN1987A data}",
    eprint = "2105.12736",
    archivePrefix = "arXiv",
    primaryClass = "hep-ph",
    reportNumber = "FERMILAB-PUB-21-225-T, NUHEP-TH/21-05, N3AS-21-009",
    doi = "10.1103/PhysRevD.105.095019",
    journal = "Phys. Rev. D",
    volume = "105",
    number = "9",
    pages = "095019",
    year = "2022"
}

@article{DeGouvea:2020ang,
    author = "De Gouv\^ea, Andr\'e and Martinez-Soler, Ivan and Perez-Gonzalez, Yuber F. and Sen, Manibrata",
    title = "{Fundamental physics with the diffuse supernova background neutrinos}",
    eprint = "2007.13748",
    archivePrefix = "arXiv",
    primaryClass = "hep-ph",
    reportNumber = "NUHEP-TH/20-08, FERMILAB-PUB-20-353-T",
    doi = "10.1103/PhysRevD.102.123012",
    journal = "Phys. Rev. D",
    volume = "102",
    pages = "123012",
    year = "2020"
}

@article{IceCube:2023ame,
    author = "Abbasi, R. and others",
    collaboration = "IceCube",
    title = "{Observation of high-energy neutrinos from the Galactic plane}",
    eprint = "2307.04427",
    archivePrefix = "arXiv",
    primaryClass = "astro-ph.HE",
    doi = "10.1126/science.adc9818",
    journal = "Science",
    volume = "380",
    number = "6652",
    pages = "adc9818",
    year = "2023"
}

@article{KM3NeT:2024paj,
    author = "Aiello, S. and others",
    collaboration = "KM3NeT",
    title = "{Astronomy potential of KM3NeT/ARCA}",
    eprint = "2402.08363",
    archivePrefix = "arXiv",
    primaryClass = "astro-ph.HE",
    doi = "10.1140/epjc/s10052-024-13137-2",
    journal = "Eur. Phys. J. C",
    volume = "84",
    number = "9",
    pages = "885",
    year = "2024"
}

@article{Arguelles:2018awr,
    author = {Arg\"uelles, Carlos A. and Palomares-Ruiz, Sergio and Schneider, Austin and Wille, Logan and Yuan, Tianlu},
    title = "{Unified atmospheric neutrino passing fractions for large-scale neutrino telescopes}",
    eprint = "1805.11003",
    archivePrefix = "arXiv",
    primaryClass = "hep-ph",
    reportNumber = "IFIC/18-24, IFIC-18-24",
    doi = "10.1088/1475-7516/2018/07/047",
    journal = "JCAP",
    volume = "07",
    pages = "047",
    year = "2018"
}

@article{deGouvea:2009fp,
    author = "de Gouv\^ea, Andre and Huang, Wei-Chih and Jenkins, James",
    title = "{Pseudo-Dirac Neutrinos in the New Standard Model}",
    eprint = "0906.1611",
    archivePrefix = "arXiv",
    primaryClass = "hep-ph",
    reportNumber = "LA-UR-09-03593, NUHEP-TH-09-08",
    doi = "10.1103/PhysRevD.80.073007",
    journal = "Phys. Rev. D",
    volume = "80",
    pages = "073007",
    year = "2009"
}

@article{Ansarifard:2022kvy,
    author = "Ansarifard, Saeed and Farzan, Yasaman",
    title = "{Revisiting pseudo-Dirac neutrino scenario after recent solar neutrino data}",
    eprint = "2211.09105",
    archivePrefix = "arXiv",
    primaryClass = "hep-ph",
    doi = "10.1103/PhysRevD.107.075029",
    journal = "Phys. Rev. D",
    volume = "107",
    number = "7",
    pages = "075029",
    year = "2023"
}

@article{Beacom:2003eu,
    author = "Beacom, John F. and Bell, Nicole F. and Hooper, Dan and Learned, John G. and Pakvasa, Sandip and Weiler, Thomas J.",
    title = "{PseudoDirac Neutrinos: A Challenge for Neutrino Telescopes}",
    eprint = "hep-ph/0307151",
    archivePrefix = "arXiv",
    reportNumber = "FERMILAB-PUB-03-201-A, MADPH-03-1337",
    doi = "10.1103/PhysRevLett.92.011101",
    journal = "Phys. Rev. Lett.",
    volume = "92",
    pages = "011101",
    year = "2004"
}

@article{Super-Kamiokande:2001ljr,
    author = "Fukuda, S. and others",
    collaboration = "Super-Kamiokande",
    title = "{Solar B-8 and hep neutrino measurements from 1258 days of Super-Kamiokande data}",
    eprint = "hep-ex/0103032",
    archivePrefix = "arXiv",
    doi = "10.1103/PhysRevLett.86.5651",
    journal = "Phys. Rev. Lett.",
    volume = "86",
    pages = "5651--5655",
    year = "2001"
}

@article{Super-Kamiokande:1998kpq,
    author = "Fukuda, Y. and others",
    collaboration = "Super-Kamiokande",
    title = "{Evidence for oscillation of atmospheric neutrinos}",
    eprint = "hep-ex/9807003",
    archivePrefix = "arXiv",
    reportNumber = "BU-98-17, ICRR-REPORT-422-98-18, UCI-98-8, KEK-PREPRINT-98-95, LSU-HEPA-5-98, UMD-98-003, SBHEP-98-5, TKU-PAP-98-06, TIT-HPE-98-09",
    doi = "10.1103/PhysRevLett.81.1562",
    journal = "Phys. Rev. Lett.",
    volume = "81",
    pages = "1562--1567",
    year = "1998"
}

@article{SNO:2002tuh,
    author = "Ahmad, Q. R. and others",
    collaboration = "SNO",
    title = "{Direct evidence for neutrino flavor transformation from neutral current interactions in the Sudbury Neutrino Observatory}",
    eprint = "nucl-ex/0204008",
    archivePrefix = "arXiv",
    doi = "10.1103/PhysRevLett.89.011301",
    journal = "Phys. Rev. Lett.",
    volume = "89",
    pages = "011301",
    year = "2002"
}

@article{Borexino:2008dzn,
    author = "Arpesella, C. and others",
    collaboration = "Borexino",
    title = "{Direct Measurement of the Be-7 Solar Neutrino Flux with 192 Days of Borexino Data}",
    eprint = "0805.3843",
    archivePrefix = "arXiv",
    primaryClass = "astro-ph",
    doi = "10.1103/PhysRevLett.101.091302",
    journal = "Phys. Rev. Lett.",
    volume = "101",
    pages = "091302",
    year = "2008"
}

@article{IceCube:2014flw,
    author = "Aartsen, M. G. and others",
    collaboration = "IceCube",
    title = "{Determining neutrino oscillation parameters from atmospheric muon neutrino disappearance with three years of IceCube DeepCore data}",
    eprint = "1410.7227",
    archivePrefix = "arXiv",
    primaryClass = "hep-ex",
    doi = "10.1103/PhysRevD.91.072004",
    journal = "Phys. Rev. D",
    volume = "91",
    number = "7",
    pages = "072004",
    year = "2015"
}

@article{KM3NeT:2024ecf,
    author = "Aiello, S. and others",
    collaboration = "KM3NeT",
    title = "{Measurement of neutrino oscillation parameters with the first six detection units of KM3NeT/ORCA}",
    eprint = "2408.07015",
    archivePrefix = "arXiv",
    primaryClass = "hep-ex",
    doi = "10.1007/JHEP10(2024)206",
    journal = "JHEP",
    volume = "10",
    pages = "206",
    year = "2024"
}

@article{Wolfenstein:1981kw,
    author = "Wolfenstein, Lincoln",
    title = "{Different Varieties of Massive Dirac Neutrinos}",
    reportNumber = "COO-3066-164",
    doi = "10.1016/0550-3213(81)90096-1",
    journal = "Nucl. Phys. B",
    volume = "186",
    pages = "147--152",
    year = "1981"
}

@article{Petcov:1982ya,
    author = "Petcov, S. T.",
    title = "{On Pseudodirac Neutrinos, Neutrino Oscillations and Neutrinoless Double beta Decay}",
    doi = "10.1016/0370-2693(82)91246-1",
    journal = "Phys. Lett. B",
    volume = "110",
    pages = "245--249",
    year = "1982"
}

@article{Valle:1983dk,
    author = "Valle, J. W. F. and Singer, M.",
    title = "{Lepton Number Violation With Quasi Dirac Neutrinos}",
    reportNumber = "RL-83-018",
    doi = "10.1103/PhysRevD.28.540",
    journal = "Phys. Rev. D",
    volume = "28",
    pages = "540",
    year = "1983"
}

@article{Kobayashi:2000md,
    author = "Kobayashi, Makoto and Lim, C. S.",
    title = "{Pseudo Dirac scenario for neutrino oscillations}",
    eprint = "hep-ph/0012266",
    archivePrefix = "arXiv",
    reportNumber = "KEK-TH-733, KOBE-TH-00-10",
    doi = "10.1103/PhysRevD.64.013003",
    journal = "Phys. Rev. D",
    volume = "64",
    pages = "013003",
    year = "2001"
}

@article{Abdullahi:2020rge,
    author = "Abdullahi, Asli and Denton, Peter B.",
    title = "{Visible Decay of Astrophysical Neutrinos at IceCube}",
    eprint = "2005.07200",
    archivePrefix = "arXiv",
    primaryClass = "hep-ph",
    doi = "10.1103/PhysRevD.102.023018",
    journal = "Phys. Rev. D",
    volume = "102",
    number = "2",
    pages = "023018",
    year = "2020"
}

@article{MacDonald:2024vtw,
    author = "MacDonald, Miller and Mart\'\i{}nez-Mirav\'e, Pablo and Tamborra, Irene",
    title = "{The Unknowns of the Diffuse Supernova Neutrino Background Hinder New Physics Searches}",
    eprint = "2409.16367",
    archivePrefix = "arXiv",
    primaryClass = "astro-ph.HE",
    doi = "10.1088/1475-7516/2025/01/062",
    journal = "JCAP",
    volume = "01",
    pages = "062",
    year = "2025"
}

@article{Beacom:2002cb,
    author = "Beacom, John F. and Bell, Nicole F.",
    title = "{Do Solar Neutrinos Decay?}",
    eprint = "hep-ph/0204111",
    archivePrefix = "arXiv",
    reportNumber = "FERMILAB-PUB-02-061-A",
    doi = "10.1103/PhysRevD.65.113009",
    journal = "Phys. Rev. D",
    volume = "65",
    pages = "113009",
    year = "2002"
}

@misc{Carloni:2025dhv,
    author = {Carloni, Kiara and Porto, Yago and Arg\"uelles, Carlos A. and Dev, P. S. Bhupal and Jana, Sudip},
    title = "{Signatures of quasi-Dirac neutrinos in diffuse high-energy astrophysical neutrino data}",
    eprint = "2503.19960",
    archivePrefix = "arXiv",
    primaryClass = "hep-ph",
    month = "3",
    year = "2025"
}

@article{Berryman:2014qha,
    author = "Berryman, Jeffrey M. and de Gouvea, Andre and Hernandez, Daniel",
    title = "{Solar Neutrinos and the Decaying Neutrino Hypothesis}",
    eprint = "1411.0308",
    archivePrefix = "arXiv",
    primaryClass = "hep-ph",
    reportNumber = "NUHEP-TH-14-08",
    doi = "10.1103/PhysRevD.92.073003",
    journal = "Phys. Rev. D",
    volume = "92",
    number = "7",
    pages = "073003",
    year = "2015"
}

@article{Valera:2024buc,
    author = "Valera, Victor B. and Fiorillo, Damiano F. G. and Esteban, Ivan and Bustamante, Mauricio",
    title = "{New limits on neutrino decay from high-energy astrophysical neutrinos}",
    eprint = "2405.14826",
    archivePrefix = "arXiv",
    primaryClass = "astro-ph.HE",
    doi = "10.1103/PhysRevD.110.043004",
    journal = "Phys. Rev. D",
    volume = "110",
    number = "4",
    pages = "043004",
    year = "2024"
}

@article{Ivanez-Ballesteros:2023lqa,
    author = "Iv{\'a}{\~n}ez-Ballesteros, Pilar and Volpe, Maria Cristina",
    title = "{SN1987A and neutrino non-radiative decay}",
    eprint = "2307.03549",
    archivePrefix = "arXiv",
    primaryClass = "hep-ph",
    doi = "10.1016/j.physletb.2023.138252",
    journal = "Phys. Lett. B",
    volume = "847",
    pages = "138252",
    year = "2023"
}

@article{Picoreti:2021yct,
    author = "Picoreti, R. and Pramanik, D. and de Holanda, P. C. and Peres, O. L. G.",
    title = "{Updating {\ensuremath{\nu}}3 lifetime from solar antineutrino spectra}",
    eprint = "2109.13272",
    archivePrefix = "arXiv",
    primaryClass = "hep-ph",
    doi = "10.1103/PhysRevD.106.015025",
    journal = "Phys. Rev. D",
    volume = "106",
    number = "1",
    pages = "015025",
    year = "2022"
}

@article{Martinez-Mirave:2024hfd,
    author = "Mart{\'\i}nez-Mirav{\'e}, Pablo and Tamborra, Irene and T{\'o}rtola, Mariam",
    title = "{The Sun and core-collapse supernovae are leading probes of the neutrino lifetime}",
    eprint = "2402.00116",
    archivePrefix = "arXiv",
    primaryClass = "astro-ph.HE",
    doi = "10.1088/1475-7516/2024/05/002",
    journal = "JCAP",
    volume = "05",
    pages = "002",
    year = "2024"
}

@article{Barenboim:2020vrr,
    author = "Barenboim, Gabriela and Chen, Joe Zhiyu and Hannestad, Steen and Oldengott, Isabel M. and Tram, Thomas and Wong, Yvonne Y. Y.",
    title = "{Invisible neutrino decay in precision cosmology}",
    eprint = "2011.01502",
    archivePrefix = "arXiv",
    primaryClass = "astro-ph.CO",
    doi = "10.1088/1475-7516/2021/03/087",
    journal = "JCAP",
    volume = "03",
    pages = "087",
    year = "2021"
}

@article{Escudero:2019gfk,
    author = "Escudero, Miguel and Fairbairn, Malcolm",
    title = "{Cosmological Constraints on Invisible Neutrino Decays Revisited}",
    eprint = "1907.05425",
    archivePrefix = "arXiv",
    primaryClass = "hep-ph",
    reportNumber = "KCL-2019-57",
    doi = "10.1103/PhysRevD.100.103531",
    journal = "Phys. Rev. D",
    volume = "100",
    number = "10",
    pages = "103531",
    year = "2019"
}

@article{Ivanez-Ballesteros:2022szu,
    author = "Ivanez-Ballesteros, Pilar and Volpe, M. Cristina",
    title = "{Neutrino nonradiative decay and the diffuse supernova neutrino background}",
    eprint = "2209.12465",
    archivePrefix = "arXiv",
    primaryClass = "hep-ph",
    doi = "10.1103/PhysRevD.107.023017",
    journal = "Phys. Rev. D",
    volume = "107",
    number = "2",
    pages = "023017",
    year = "2023"
}

@article{P-ONE:2020ljt,
    author = "Agostini, Matteo and others",
    collaboration = "P-ONE",
    title = "{The Pacific Ocean Neutrino Experiment}",
    eprint = "2005.09493",
    archivePrefix = "arXiv",
    primaryClass = "astro-ph.HE",
    doi = "10.1038/s41550-020-1182-4",
    journal = "Nature Astron.",
    volume = "4",
    number = "10",
    pages = "913--915",
    year = "2020"
}

@article{Baikal-GVD:2018isr,
    author = "Avrorin, A. D. and others",
    editor = "Volkova, V. E. and Zhezher, Y. V. and Levkov, D. G. and Rubakov, V. A. and Matveev, V. A.",
    collaboration = "Baikal-GVD",
    title = "{Baikal-GVD: status and prospects}",
    eprint = "1808.10353",
    archivePrefix = "arXiv",
    primaryClass = "astro-ph.IM",
    doi = "10.1051/epjconf/201819101006",
    journal = "EPJ Web Conf.",
    volume = "191",
    pages = "01006",
    year = "2018"
}

@article{Johannesson:2018bit,
    author = "J{\'o}hannesson, Gu{\dj}laugur and Porter, Troy A. and Moskalenko, Igor V.",
    title = "{The Three-Dimensional Spatial Distribution of Interstellar Gas in the Milky Way: Implications for Cosmic Rays and High-Energy Gamma-Ray Emissions}",
    eprint = "1802.08646",
    archivePrefix = "arXiv",
    primaryClass = "astro-ph.HE",
    doi = "10.3847/1538-4357/aab26e",
    journal = "Astrophys. J.",
    volume = "856",
    number = "1",
    pages = "45",
    year = "2018"
}

@article{Soding:2025scd,
   title={Spatially coherent 3D distributions of HI and CO in the Milky Way},
   volume={693},
   ISSN={1432-0746},
   url={http://dx.doi.org/10.1051/0004-6361/202451361},
   DOI={10.1051/0004-6361/202451361},
   journal={Astronomy \& Astrophysics},
   publisher={EDP Sciences},
   author={Söding, Laurin and Edenhofer, Gordian and Enßlin, Torsten A. and Frank, Philipp and Kissmann, Ralf and Phan, Vo Hong Minh and Ramírez, Andrés and Zandinejad, Hanieh and Mertsch, Philipp},
   year={2025},
   month=jan, pages={A139} }

@article{Fermi-LAT:2012edv,
    author = "Ackermann, M. and others",
    collaboration = "Fermi-LAT",
    title = "{Fermi-LAT Observations of the Diffuse Gamma-Ray Emission: Implications for Cosmic Rays and the Interstellar Medium}",
    eprint = "1202.4039",
    archivePrefix = "arXiv",
    primaryClass = "astro-ph.HE",
    doi = "10.1088/0004-637X/750/1/3",
    journal = "Astrophys. J.",
    volume = "750",
    pages = "3",
    year = "2012"
}

@article{KATRIN:2024cdt,
    author = "Aker, Max and others",
    collaboration = "KATRIN",
    title = "{Direct neutrino-mass measurement based on 259 days of KATRIN data}",
    eprint = "2406.13516",
    archivePrefix = "arXiv",
    primaryClass = "nucl-ex",
    doi = "10.1126/science.adq9592",
    journal = "Science",
    volume = "388",
    number = "6743",
    pages = "adq9592",
    year = "2025"
}

@article{Bahcall:1972my,
    author = "Bahcall, John N. and Cabibbo, N. and Yahil, A.",
    title = "{Are neutrinos stable particles?}",
    doi = "10.1103/PhysRevLett.28.316",
    journal = "Phys. Rev. Lett.",
    volume = "28",
    pages = "316--318",
    year = "1972"
}

@article{Bahcall:1986gq,
    author = "Bahcall, John N. and Petcov, S. T. and Toshev, S. and Valle, J. W. F.",
    title = "{Tests of Neutrino Stability}",
    reportNumber = "PRINT-86-1185 (IAS,PRINCETON)",
    doi = "10.1016/0370-2693(86)90065-1",
    journal = "Phys. Lett. B",
    volume = "181",
    pages = "369--374",
    year = "1986"
}

@article{Nussinov:1987pc,
    author = "Nussinov, S.",
    title = "{Some Comments on Decaying Neutrinos and the Triplet Majoron Model}",
    doi = "10.1016/0370-2693(87)91548-6",
    journal = "Phys. Lett. B",
    volume = "185",
    pages = "171--176",
    year = "1987"
}

@article{Kim:1990km,
    author = "Kim, C. W. and Lam, W. P.",
    title = "{Some remarks on neutrino decay via a Nambu-Goldstone boson}",
    doi = "10.1142/S0217732390000354",
    journal = "Mod. Phys. Lett. A",
    volume = "5",
    pages = "297--299",
    year = "1990"
}

@article{Chikashige:1980qk,
    author = "Chikashige, Y. and Mohapatra, Rabindra N. and Peccei, R. D.",
    title = "{Spontaneously Broken Lepton Number and Cosmological Constraints on the Neutrino Mass Spectrum}",
    reportNumber = "MPI-PAE/PTh 40/80",
    doi = "10.1103/PhysRevLett.45.1926",
    journal = "Phys. Rev. Lett.",
    volume = "45",
    pages = "1926",
    year = "1980"
}

@article{Gelmini:1980re,
    author = "Gelmini, G. B. and Roncadelli, M.",
    title = "{Left-Handed Neutrino Mass Scale and Spontaneously Broken Lepton Number}",
    reportNumber = "MPI-PAE-PTH-50-80",
    doi = "10.1016/0370-2693(81)90559-1",
    journal = "Phys. Lett. B",
    volume = "99",
    pages = "411--415",
    year = "1981"
}

@article{Gelmini:1983ea,
    author = "Gelmini, G. B. and Valle, J. W. F.",
    title = "{Fast Invisible Neutrino Decays}",
    reportNumber = "CERN-TH-3791",
    doi = "10.1016/0370-2693(84)91258-9",
    journal = "Phys. Lett. B",
    volume = "142",
    pages = "181--187",
    year = "1984"
}

@article{Shrock:1974nd,
    author = "Shrock, R.",
    title = "{Decay l0 ---\ensuremath{>} nu(lepton) gamma in gauge theories of weak and electromagnetic interactions}",
    doi = "10.1103/PhysRevD.9.743",
    journal = "Phys. Rev. D",
    volume = "9",
    pages = "743--748",
    year = "1974"
}

@article{Marciano:1977wx,
    author = "Marciano, W. J. and Sanda, A. I.",
    title = "{Exotic Decays of the Muon and Heavy Leptons in Gauge Theories}",
    reportNumber = "COO-2232B-116",
    doi = "10.1016/0370-2693(77)90377-X",
    journal = "Phys. Lett. B",
    volume = "67",
    pages = "303--305",
    year = "1977"
}

@article{Petcov:1976ff,
    author = "Petcov, S. T.",
    title = "{The Processes $\mu \rightarrow e + \gamma, \mu \rightarrow e + \overline{e}, \nu' \rightarrow \nu + \gamma$ in the Weinberg-Salam Model with Neutrino Mixing}",
    reportNumber = "JINR-E2-10176, JINR-P2-9595",
    journal = "Sov. J. Nucl. Phys.",
    volume = "25",
    pages = "340",
    year = "1977",
    note = "[Erratum: Sov.J.Nucl.Phys. 25, 698 (1977), Erratum: Yad.Fiz. 25, 1336 (1977)]"
}

@article{Shrock:1982sc,
    author = "Shrock, Robert E.",
    title = "{Electromagnetic Properties and Decays of Dirac and Majorana Neutrinos in a General Class of Gauge Theories}",
    reportNumber = "ITP-SB-82-2",
    doi = "10.1016/0550-3213(82)90273-5",
    journal = "Nucl. Phys. B",
    volume = "206",
    pages = "359--379",
    year = "1982"
}

@article{Pal:1981rm,
    author = "Pal, Palash B. and Wolfenstein, Lincoln",
    title = "{Radiative Decays of Massive Neutrinos}",
    reportNumber = "COO-3066-167-REV, COO-3066-167",
    doi = "10.1103/PhysRevD.25.766",
    journal = "Phys. Rev. D",
    volume = "25",
    pages = "766",
    year = "1982"
}

@article{Zatsepin:1978iy,
    author = "Zatsepin, G. T. and Smirnov, A. Yu.",
    title = "{Neutrino Decay in Gauge Theories}",
    journal = "Yad. Fiz.",
    volume = "28",
    pages = "1569--1579",
    year = "1978"
}

@article{Frieman:1987as,
    author = "Frieman, Joshua A. and Haber, Howard E. and Freese, Katherine",
    title = "{Neutrino Mixing, Decays and Supernova Sn1987a}",
    reportNumber = "SLAC-PUB-4261, SCIPP-87-90, NSF-ITP-87-53",
    doi = "10.1016/0370-2693(88)91120-3",
    journal = "Phys. Lett. B",
    volume = "200",
    pages = "115--121",
    year = "1988"
}

@article{Schechter:1981cv,
    author = "Schechter, J. and Valle, J. W. F.",
    title = "{Neutrino Decay and Spontaneous Violation of Lepton Number}",
    reportNumber = "SU-4217-203, COO-3533-203",
    doi = "10.1103/PhysRevD.25.774",
    journal = "Phys. Rev. D",
    volume = "25",
    pages = "774",
    year = "1982"
}

@article{CRPropa:2022ovg,
    author = "Alves Batista, Rafael and others",
    collaboration = "CRPropa",
    title = "{CRPropa 3.2 {\textemdash} an advanced framework for high-energy particle propagation in extragalactic and galactic spaces}",
    eprint = "2208.00107",
    archivePrefix = "arXiv",
    primaryClass = "astro-ph.HE",
    doi = "10.1088/1475-7516/2022/09/035",
    journal = "JCAP",
    volume = "09",
    pages = "035",
    year = "2022"
}

@misc{Carloni:2025upc,
    author = "Carloni, Kiara and Arguelles, Carlos A and MacDonald, Miller and Mart\'\i{}nez-Soler, Ivan and Alves Batista, Rafael",
    title = "{TANDEM: A New Model of Galactic Neutrino Emission Using CR-Propa}",
    note="Upcoming, 2026."
}

@article{Esteban:2024eli,
    author = "Esteban, Ivan and Gonzalez-Garcia, M. C. and Maltoni, Michele and Martinez-Soler, Ivan and Pinheiro, Jo{\~a}o Paulo and Schwetz, Thomas",
    title = "{NuFit-6.0: updated global analysis of three-flavor neutrino oscillations}",
    eprint = "2410.05380",
    archivePrefix = "arXiv",
    primaryClass = "hep-ph",
    reportNumber = "IFT-UAM/CSIC-24-140, YITP-SB-2024-24, IPPP/24/64, IPPP/24/64, IFT-UAM/CSIC-24-140, YITP-SB-2024-24",
    doi = "10.1007/JHEP12(2024)216",
    journal = "JHEP",
    volume = "12",
    pages = "216",
    year = "2024"
}

@misc{DeLaTorreLuque:2025zsv,
    author = "De La Torre Luque, Pedro and Gaggero, Daniele and Grasso, Dario and Marinelli, Antonio and Rocamora, Manuel",
    title = "{The cosmic-ray sea explains the diffuse Galactic gamma-ray and neutrino emission from GeV to PeV}",
    eprint = "2502.18268",
    archivePrefix = "arXiv",
    primaryClass = "astro-ph.HE",
    month = "2",
    year = "2025"
}

@article{Schwefer:2022zly,
    author = "Schwefer, Georg and Mertsch, Philipp and Wiebusch, Christopher",
    title = "{Diffuse Emission of Galactic High-energy Neutrinos from a Global Fit of Cosmic Rays}",
    eprint = "2211.15607",
    archivePrefix = "arXiv",
    primaryClass = "astro-ph.HE",
    reportNumber = "TTK-22-40",
    doi = "10.3847/1538-4357/acc1e2",
    journal = "Astrophys. J.",
    volume = "949",
    number = "1",
    pages = "16",
    year = "2023"
}

@misc{Marinos:2025ddy,
    author = "Marinos, P. D. and Porter, T. A. and Rowell, G. P. and Moskalenko, I. V. and J{\'o}hannesson, G.",
    title = "{Simulating the Diffuse Neutrino Emission from the Milky Way with GALPROP}",
    eprint = "2511.09777",
    archivePrefix = "arXiv",
    primaryClass = "astro-ph.HE",
    month = "11",
    year = "2025"
}

@article{Fang:2023azx,
    author = "Fang, Ke and Gallagher, John S. and Halzen, Francis",
    title = "{The Milky Way revealed to be a neutrino desert by the IceCube Galactic plane observation}",
    eprint = "2306.17275",
    archivePrefix = "arXiv",
    primaryClass = "astro-ph.HE",
    doi = "10.1038/s41550-023-02128-0",
    journal = "Nature Astron.",
    volume = "8",
    number = "2",
    pages = "241--246",
    year = "2024"
}

@article{GonzalezGarcia:2008idc,
   title={Status of oscillation plus decay of atmospheric and long-baseline neutrinos},
   volume={663},
   ISSN={0370-2693},
   url={http://dx.doi.org/10.1016/j.physletb.2008.04.041},
   DOI={10.1016/j.physletb.2008.04.041},
   number={5},
   journal={Physics Letters B},
   publisher={Elsevier BV},
   author={Gonzalez-Garcia, M.C. and Maltoni, Michele},
   year={2008},
   month=jun, pages={405–409} }

@article{Unger:2024cmf,
   title={The Coherent Magnetic Field of the Milky Way},
   volume={970},
   ISSN={1538-4357},
   url={http://dx.doi.org/10.3847/1538-4357/ad4a54},
   DOI={10.3847/1538-4357/ad4a54},
   number={1},
   journal={The Astrophysical Journal},
   publisher={American Astronomical Society},
   author={Unger, Michael and Farrar, Glennys R.},
   year={2024},
   month=jul, pages={95} }

@article{Leal:2025eou,
    author = "Leal, Luighi P. S. and Naredo-Tuero, Daniel and Funchal, Renata Zukanovich",
    title = "{Cosmogenic neutrinos as probes of new physics}",
    eprint = "2504.10576",
    archivePrefix = "arXiv",
    primaryClass = "hep-ph",
    reportNumber = "IFT-UAM/CSIC-25-38",
    doi = "10.1007/JHEP08(2025)057",
    journal = "JHEP",
    volume = "08",
    pages = "057",
    year = "2025"
}

@article{Dembinski:2017zsh,
    author = "Dembinski, Hans Peter and Engel, Ralph and Fedynitch, Anatoli and Gaisser, Thomas and Riehn, Felix and Stanev, Todor",
    title = "{Data-driven model of the cosmic-ray flux and mass composition from 10 GeV to $10^{11}$ GeV}",
    eprint = "1711.11432",
    archivePrefix = "arXiv",
    primaryClass = "astro-ph.HE",
    doi = "10.22323/1.301.0533",
    journal = "PoS",
    volume = "ICRC2017",
    pages = "533",
    year = "2018"
}

@article{TRIDENT:2022hql,
    author = "Ye, Z. P. and others",
    collaboration = "TRIDENT",
    title = "{A multi-cubic-kilometre neutrino telescope in the western Pacific Ocean}",
    eprint = "2207.04519",
    archivePrefix = "arXiv",
    primaryClass = "astro-ph.HE",
    doi = "10.1038/s41550-023-02087-6",
    journal = "Nature Astron.",
    volume = "7",
    number = "12",
    pages = "1497--1505",
    year = "2023"
}

@article{Huang:2023mzt,
    author = "Huang, Tian-Qi and Cao, Zhen and Chen, Mingjun and Liu, Jiali and Wang, Zike and You, Xiaohao and Qi, Ying",
    title = "{Proposal for the High Energy Neutrino Telescope}",
    doi = "10.22323/1.444.1080",
    journal = "PoS",
    volume = "ICRC2023",
    pages = "1080",
    year = "2023"
}

@article{Mertsch:2021gco,
   title={{Bayesian inference of three-dimensional gas maps: I. Galactic CO}},
   volume={655},
   ISSN={1432-0746},
   url={http://dx.doi.org/10.1051/0004-6361/202141000},
   DOI={10.1051/0004-6361/202141000},
   journal={Astronomy \& Astrophysics},
   publisher={EDP Sciences},
   author={Mertsch, P. and Vittino, A.},
   year={2021},
   month=nov, pages={A64} }

@article{Mertsch:2023ghi,
   title={{Bayesian inference of three-dimensional gas maps: II. Galactic HI}},
   volume={671},
   ISSN={1432-0746},
   url={http://dx.doi.org/10.1051/0004-6361/202243326},
   DOI={10.1051/0004-6361/202243326},
   journal={Astronomy \& Astrophysics},
   publisher={EDP Sciences},
   author={Mertsch, P. and Phan, V. H. M.},
   year={2023},
   month=mar, pages={A54} }

@article{Bertolini:1987kz,
    author = "Bertolini, S. and Santamaria, A.",
    title = "{The Doublet Majoron Model and Solar Neutrino Oscillations}",
    reportNumber = "CMU-HEP87-31",
    doi = "10.1016/0550-3213(88)90100-9",
    journal = "Nucl. Phys. B",
    volume = "310",
    pages = "714--742",
    year = "1988"
}

@article{Fuller:1988ega,
    author = "Fuller, G. M. and Mayle, R. and Wilson, J. R.",
    title = "{The Majoron model and stellar collapse}",
    doi = "10.1086/166695",
    journal = "Astrophys. J.",
    volume = "332",
    pages = "826",
    year = "1988"
}

@article{Franklin:2023diy,
    author = "Franklin, Jack and Perez-Gonzalez, Yuber F. and Turner, Jessica",
    title = "{JUNO as a probe of the pseudo-Dirac nature using solar neutrinos}",
    eprint = "2304.05418",
    archivePrefix = "arXiv",
    primaryClass = "hep-ph",
    reportNumber = "IPPP/23/20",
    doi = "10.1103/PhysRevD.108.035010",
    journal = "Phys. Rev. D",
    volume = "108",
    number = "3",
    pages = "035010",
    year = "2023"
}

@article{LHAASO:2021gok,
    author = "Cao, Zhen and others",
    collaboration = "LHAASO",
    title = "{Ultrahigh-energy photons up to 1.4 petaelectronvolts from 12 $\gamma$-ray Galactic sources}",
    doi = "10.1038/s41586-021-03498-z",
    journal = "Nature",
    volume = "594",
    number = "7861",
    pages = "33--36",
    year = "2021"
}

@article{HESS:2016pst,
    author = "Abramowski, A. and others",
    collaboration = "H.E.S.S.",
    title = "{Acceleration of petaelectronvolt protons in the Galactic Centre}",
    eprint = "1603.07730",
    archivePrefix = "arXiv",
    primaryClass = "astro-ph.HE",
    doi = "10.1038/nature17147",
    journal = "Nature",
    volume = "531",
    pages = "476",
    year = "2016"
}

@article{Margiotta:2022kid,
    author = "Margiotta, Annarita",
    collaboration = "KM3NeT",
    title = "{The KM3NeT infrastructure: Status and first results}",
    eprint = "2208.07370",
    archivePrefix = "arXiv",
    primaryClass = "astro-ph.IM",
    doi = "10.21468/SciPostPhysProc.13.030",
    journal = "SciPost Phys. Proc.",
    volume = "13",
    pages = "030",
    year = "2023"
}

@article{IceCube:2020wum,
    author = "Abbasi, R. and others",
    collaboration = "IceCube",
    title = "{The IceCube high-energy starting event sample: Description and flux characterization with 7.5 years of data}",
    eprint = "2011.03545",
    archivePrefix = "arXiv",
    primaryClass = "astro-ph.HE",
    doi = "10.1103/PhysRevD.104.022002",
    journal = "Phys. Rev. D",
    volume = "104",
    pages = "022002",
    year = "2021"
}

@article{KM3Net:2016zxf,
    author = "Adrian-Martinez, S. and others",
    collaboration = "KM3Net",
    title = "{Letter of intent for KM3NeT 2.0}",
    eprint = "1601.07459",
    archivePrefix = "arXiv",
    primaryClass = "astro-ph.IM",
    doi = "10.1088/0954-3899/43/8/084001",
    journal = "J. Phys. G",
    volume = "43",
    number = "8",
    pages = "084001",
    year = "2016"
}

@article{Koldobskiy:2021nld,
    author = "Koldobskiy, S. and Kachelrie{\ss}, M. and Lskavyan, A. and Neronov, A. and Ostapchenko, S. and Semikoz, D. V.",
    title = "{Energy spectra of secondaries in proton-proton interactions}",
    eprint = "2110.00496",
    archivePrefix = "arXiv",
    primaryClass = "astro-ph.HE",
    doi = "10.1103/PhysRevD.104.123027",
    journal = "Phys. Rev. D",
    volume = "104",
    number = "12",
    pages = "123027",
    year = "2021"
}

@article{Kachelriess:2019ifk,
    author = "Kachelrie{\ss}, M. and Moskalenko, I. V. and Ostapchenko, S.",
    title = "{AAfrag: Interpolation routines for Monte Carlo results on secondary production in proton-proton, proton-nucleus and nucleus-nucleus interactions}",
    eprint = "1904.05129",
    archivePrefix = "arXiv",
    primaryClass = "hep-ph",
    doi = "10.1016/j.cpc.2019.08.001",
    journal = "Comput. Phys. Commun.",
    volume = "245",
    pages = "106846",
    year = "2019"
}

@article{Silagadze:1995tr,
    author = "Silagadze, Z. K.",
    title = "{Neutrino mass and the mirror universe}",
    eprint = "hep-ph/9503481",
    archivePrefix = "arXiv",
    reportNumber = "JINR-E2-95-150",
    journal = "Phys. Atom. Nucl.",
    volume = "60",
    pages = "272--275",
    year = "1997"
}

@article{Joshipura:2013yba,
    author = "Joshipura, Anjan S. and Mohanty, Subhendra and Pakvasa, Sandip",
    title = "{Pseudo-Dirac neutrinos via a mirror world and depletion of ultrahigh energy neutrinos}",
    eprint = "1307.5712",
    archivePrefix = "arXiv",
    primaryClass = "hep-ph",
    doi = "10.1103/PhysRevD.89.033003",
    journal = "Phys. Rev. D",
    volume = "89",
    number = "3",
    pages = "033003",
    year = "2014"
}

@article{Gu:2006dc,
    author = "Gu, Pei-Hong and He, Hong-Jian",
    title = "{Neutrino Mass and Baryon Asymmetry from Dirac Seesaw}",
    eprint = "hep-ph/0610275",
    archivePrefix = "arXiv",
    doi = "10.1088/1475-7516/2006/12/010",
    journal = "JCAP",
    volume = "12",
    pages = "010",
    year = "2006"
}

@article{Ma:2014qra,
    author = "Ma, Ernest and Srivastava, Rahul",
    title = "{Dirac or inverse seesaw neutrino masses with $B-L$ gauge symmetry and $S_3$ flavor symmetry}",
    eprint = "1411.5042",
    archivePrefix = "arXiv",
    primaryClass = "hep-ph",
    doi = "10.1016/j.physletb.2014.12.049",
    journal = "Phys. Lett. B",
    volume = "741",
    pages = "217--222",
    year = "2015"
}

@article{Valle:2016kyz,
    author = "Valle, Jos\'e W. F. and Vaquera-Araujo, C. A.",
    title = "{Dynamical seesaw mechanism for Dirac neutrinos}",
    eprint = "1601.05237",
    archivePrefix = "arXiv",
    primaryClass = "hep-ph",
    reportNumber = "IFIC-16-XX",
    doi = "10.1016/j.physletb.2016.02.031",
    journal = "Phys. Lett. B",
    volume = "755",
    pages = "363--366",
    year = "2016"
}

@article{CentellesChulia:2018bkz,
    author = "Centelles Chuli\'a, Salvador and Srivastava, Rahul and Valle, Jos\'e W. F.",
    title = "{Seesaw Dirac neutrino mass through dimension-six operators}",
    eprint = "1804.03181",
    archivePrefix = "arXiv",
    primaryClass = "hep-ph",
    reportNumber = "IFIC/18-xxx, IFIC-18-XXX",
    doi = "10.1103/PhysRevD.98.035009",
    journal = "Phys. Rev. D",
    volume = "98",
    number = "3",
    pages = "035009",
    year = "2018"
}

@article{Lee:1956qn,
    author = "Lee, T. D. and Yang, Chen-Ning",
    title = "{Question of Parity Conservation in Weak Interactions}",
    doi = "10.1103/PhysRev.104.254",
    journal = "Phys. Rev.",
    volume = "104",
    pages = "254--258",
    year = "1956"
}

@article{Foot:1991py,
    author = "Foot, Robert and Lew, H. and Volkas, R. R.",
    title = "{Possible consequences of parity conservation}",
    reportNumber = "SHEP-90-91-36, UM-P-91-82, OZ-91-11",
    doi = "10.1142/S0217732392004031",
    journal = "Mod. Phys. Lett. A",
    volume = "7",
    pages = "2567--2574",
    year = "1992"
}

@article{Berezhiani:1995yi,
    author = "Berezhiani, Zurab G. and Mohapatra, Rabindra N.",
    title = "{Reconciling present neutrino puzzles: Sterile neutrinos as mirror neutrinos}",
    eprint = "hep-ph/9505385",
    archivePrefix = "arXiv",
    reportNumber = "INFN-FE-06-95",
    doi = "10.1103/PhysRevD.52.6607",
    journal = "Phys. Rev. D",
    volume = "52",
    pages = "6607--6611",
    year = "1995"
}

@article{Jansson:2012rt,
    author = "Jansson, Ronnie and Farrar, Glennys R.",
    title = "{The Galactic Magnetic Field}",
    eprint = "1210.7820",
    archivePrefix = "arXiv",
    primaryClass = "astro-ph.GA",
    doi = "10.1088/2041-8205/761/1/L11",
    journal = "Astrophys. J. Lett.",
    volume = "761",
    pages = "L11",
    year = "2012"
}

@article{IceCube:2025tgp,
    author = "Abbasi, R. and others",
    collaboration = "IceCube",
    title = "{Evidence for a Spectral Break or Curvature in the Spectrum of Astrophysical Neutrinos from 5 TeV--10 PeV}",
    eprint = "2507.22233",
    archivePrefix = "arXiv",
    primaryClass = "astro-ph.HE",
    doi = "10.1103/2gh9-d4q7",
    journal = "Phys. Rev. Lett.",
    volume = "136",
    pages = "121002",
    year = "2026"
}

@article{Chen:2022idm,
    author = "Chen, Joe Zhiyu and Oldengott, Isabel M. and Pierobon, Giovanni and Wong, Yvonne Y. Y.",
    title = "{Weaker yet again: mass spectrum-consistent cosmological constraints on the neutrino lifetime}",
    eprint = "2203.09075",
    archivePrefix = "arXiv",
    primaryClass = "hep-ph",
    doi = "10.1140/epjc/s10052-022-10518-3",
    journal = "Eur. Phys. J. C",
    volume = "82",
    number = "7",
    pages = "640",
    year = "2022"
}

\clearpage
\pagebreak
\appendix
\onecolumngrid
\ifx \standalonesupplemental\undefined
\setcounter{page}{1}
\counterwithin{figure}{section}
\numberwithin{equation}{section}
\fi
\renewcommand{\thepage}{Supplemental Methods and Tables --- S\arabic{page}}
\renewcommand{\figurename}{SUPPL. FIG.}
\renewcommand{\tablename}{SUPPL. TABLE}
\renewcommand{\theequation}{A\arabic{equation}}
\clearpage
\begin{center}
\textbf{\large Supplemental Material}
\end{center}


\section{Formalism for visible neutrino decay\label{sec:appvisdecay}}

We consider two-body visible neutrino decay, $\nu_j \to \nu_i + X$. 
Visible neutrino decay scenarios have a rich phenomenology~\cite{Abdullahi:2020rge, MacDonald:2024vtw}, which in general depends on currently unknown neutrino properties like the absolute neutrino mass scale and the neutrino mass ordering. 
We assume normal neutrino mass ordering with only one unstable neutrino mass state, the heaviest mass state $\nu_3$, which decays to the lightest neutrino mass state $\nu_1$. 
We thus expect that the characteristic signature of this scenario is a dip in the $\nu_3$ spectrum along with a bump in the $\nu_1$ spectrum. 
The mass state fluxes read~\cite{Martinez-Soler:2021unz}
\begin{equation}
\label{eq:visibledecays}
    \begin{aligned}
        \phi_{\nu_3}(E_\nu, r, \ell, b) &= \int_r^\infty dr'~\sum_\beta |U_{\beta 3}|^2F_\beta(E_\nu, r', \ell, b)\exp\left\{ -\frac{\alpha_3 L}{E_\nu} \right\}, \\
        \phi_{\nu_2}(E_\nu, r,\ell, b) &= \int_r^\infty dr'~\sum_\beta |U_{\beta 2}|^2F_\beta(E_\nu, r', \ell, b), \\
        \phi_{\nu_1}(E_\nu, r,\ell, b) &= \int_r^\infty dr' \Biggl\{~\sum_\beta |U_{\beta 1}|^2F_\beta(E_\nu, r', \ell, b) \\
        &\qquad\qquad + \int_{E_\nu}^\infty dE'_\nu~\phi_{\nu_3}(E'_\nu, r', \ell, b) \frac{\alpha_3 B(\nu_3 \to \nu_1)}{E'_\nu} \psi_{\nu_3 \to \nu_1}(E'_\nu, E_\nu) \\
        &\qquad\qquad + \int_{E_\nu}^\infty dE'_\nu~\phi_{\bar{\nu}_3}(E'_\nu, r', \ell, b) \frac{\alpha_3 B(\bar{\nu}_3 \to \nu_1)}{E'_\nu} \psi_{\bar{\nu}_3 \to \nu_1}(E'_\nu, E_\nu) \Biggr\}.
        \end{aligned}
\end{equation}
Here, the integrals in energy represent contributions at a given $r$ along the line of sight to the $\nu_1$ flux from decay products of the $\nu_3 \to \nu_1 + X$ channel further out along the line of sight. $B(\nu_3 \to \nu_1)$ and $B(\bar{\nu}_3 \to \nu_1)$ correspond to the branching ratios between the heavier (anti)neutrino mass state and the neutrino decay product. $\psi_{\nu_3 \to \nu_1}(E_\nu', E_\nu)$ and $\psi_{\bar{\nu}_3 \to \nu_1}(E_\nu', E_\nu)$ are the normalized decay energy spectra. Analogous expressions can be written for $\bar{\nu}_i$ flux.

In general, the branching ratios and decay energy spectra could be dependent on the absolute neutrino mass scale. In this study, we only consider the \textit{quasi-degenerate} limit, which corresponds to a large absolute mass scale and a case where the neutrino masses are almost equal; more specifically, for each pair of mass states $m_l$ and $m_h$, we have $m_h \simeq m_l \gg m_h - m_l$. This limit is still applicable to the upper region of mass parameter space allowed by current direct mass constraints~\cite{KATRIN:2024cdt}. In this limiting case, the decay product energy spectrum reduces to a delta function
\begin{equation}
    \psi_{\nu_3 \to \nu_1}(E_\nu', E_\nu) = \delta(E_\nu' - E_\nu),
\end{equation}
collapsing the energy integrals in \eqref{eq:visibledecays}~\cite{Beacom:2002cb}. A more complete treatment of visible decay that leaves the absolute mass scale as a free parameter is outside the scope of this work.

We evaluate the flux predictions in the flavor basis at Earth by taking 
\begin{equation}
    \phi_{\nu_\alpha}(E_\nu, \ell, b) = \sum_{i} |U_{\alpha i}|^2 \phi_{\nu_i}(E_\nu, r = 0, \ell, b).
\end{equation}

In certain decay models, like the Majoron models considered in~\cite{Picoreti:2021yct}, the quasi-degenerate limit will sometimes also lead to suppression of helicity-violating decays, i.e. $\nu_3 \to \bar{\nu}_1$. This occurs if the Majorons couple to neutrinos through only scalar interactions. These considerations are relevant for detection channels sensitive to neutrino type, like the solar neutrinos considered in their study. The analysis in~\cite{Picoreti:2021yct} places bounds based on solar $\bar{\nu}_e$ nondetection, and thus loses sensitivity to the scalar interaction scenario in the quasi-degenerate limit. However, because IceCube and KM3NeT detect both $\nu$ and $\bar{\nu}$ and are (largely) insensitive to $\nu/\bar{\nu}$, with the sensitivity coming only from flavor and spectral features, our sensitivities apply more broadly to Majoron models with suppressed helicity-violating decay channels.

\section{Comparison of BSM phenomenology and constraints across the \texttt{TANDEM} model suite\label{sec:appdiffgasmodels}}

\begin{figure*}
  \includegraphics[width=\textwidth]{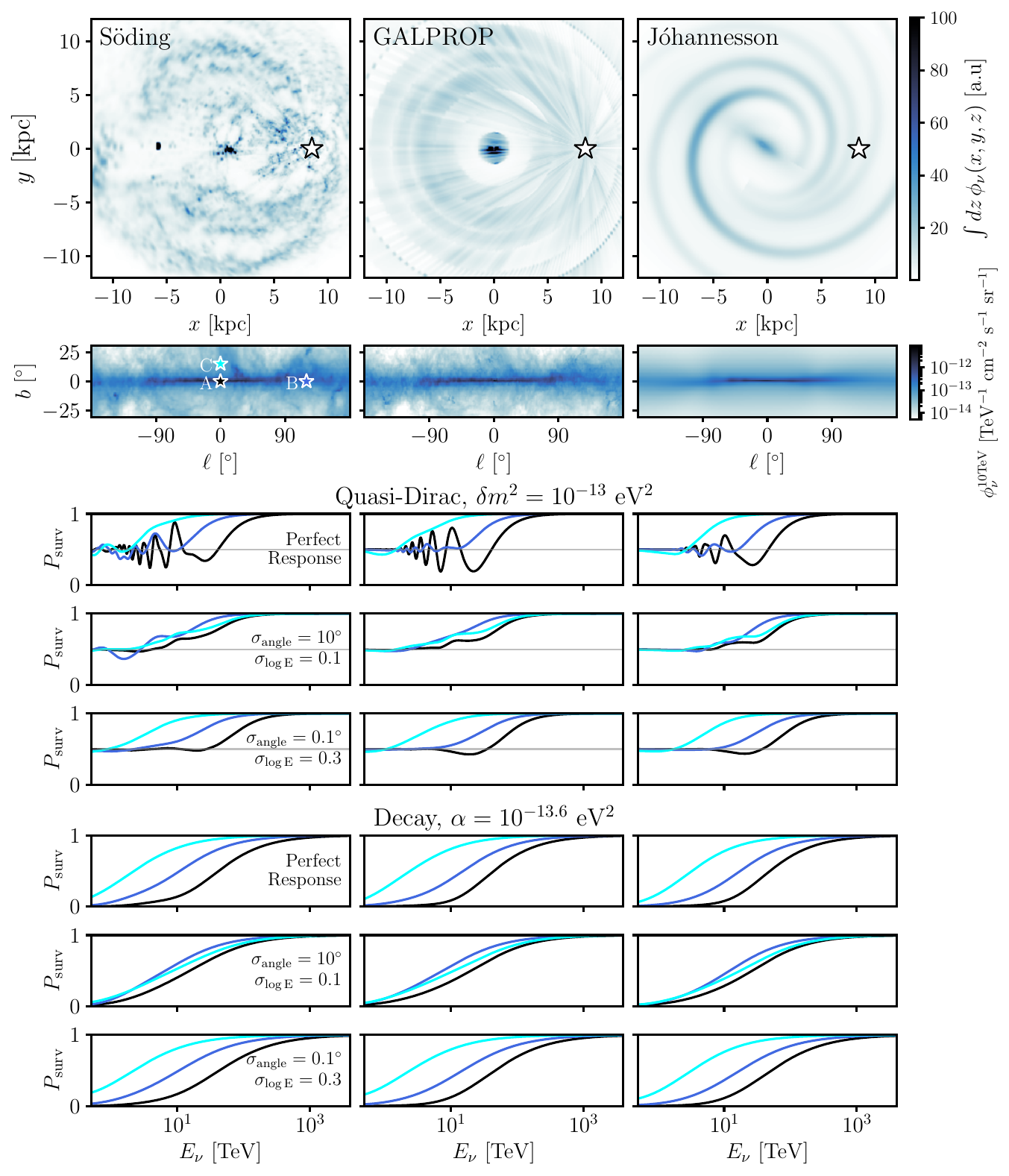}
  \caption[]{\textbf{\textit{Survival probability of neutrino emission along different lines of sight in a quasi-Dirac and decay scenarios: comparison across gas maps.}} \textit{Row 1:} Maps of $z$-integrated Galactic neutrino emission at $10~\mathrm{TeV}$ under a SM scenario assuming that Galactic gas follows the models presented in Refs.~\cite{Soding:2025scd} (left), \cite{Fermi-LAT:2012edv} (middle), and \cite{Johannesson:2018bit} (right). \textit{Row 2:} Galactic neutrino flux under each gas model assumption as a function of Galactic latitude $\ell$ and longitude $b$. The same representative sky locations as \Cref{fig:oscsmears_QD_soding} are marked. \textit{Rows 3-5:} Same as the bottom three rows of \Cref{fig:oscsmears_QD_soding}, but shown for the corresponding gas maps in each column. \textit{Rows 6-8:} Same as \Cref{fig:oscsmears_decay_soding}, but shown for the corresponding gas maps in each column.}
  \label{fig:oscsmears_QDdecay_summary}
\end{figure*}

In this section we describe how the QD and neutrino decay phenomenology varies with different \texttt{TANDEM} models of the neutrino emission.
In Fig.~\ref{fig:oscsmears_QDdecay_summary}, we show survival probabilities for QD and neutrino decay scenarios produced using \texttt{TANDEM} models based on different gas maps.
As is evident in the first and second rows of the figure, the \texttt{TANDEM} emission profile varies significantly depending on the gas model used, both in direction and in baseline distribution along the same line of sight.
However, despite differences in the 3D emission, the intrinsic smearing of the ultra-long baseline propagation effects caused by the baseline dependent emission yields broadly similar phenomenology across all three gas maps, as is shown in rows 3 (QD) and 6 (decays) of Fig. \ref{fig:oscsmears_QDdecay_summary}. 
After convolving with angular and energy resolutions characteristic to IceCube and KM3NeT, the extant differences wash out even more (rows 4-5, 7-8). 

\begin{figure*}
  \includegraphics[width=\textwidth]{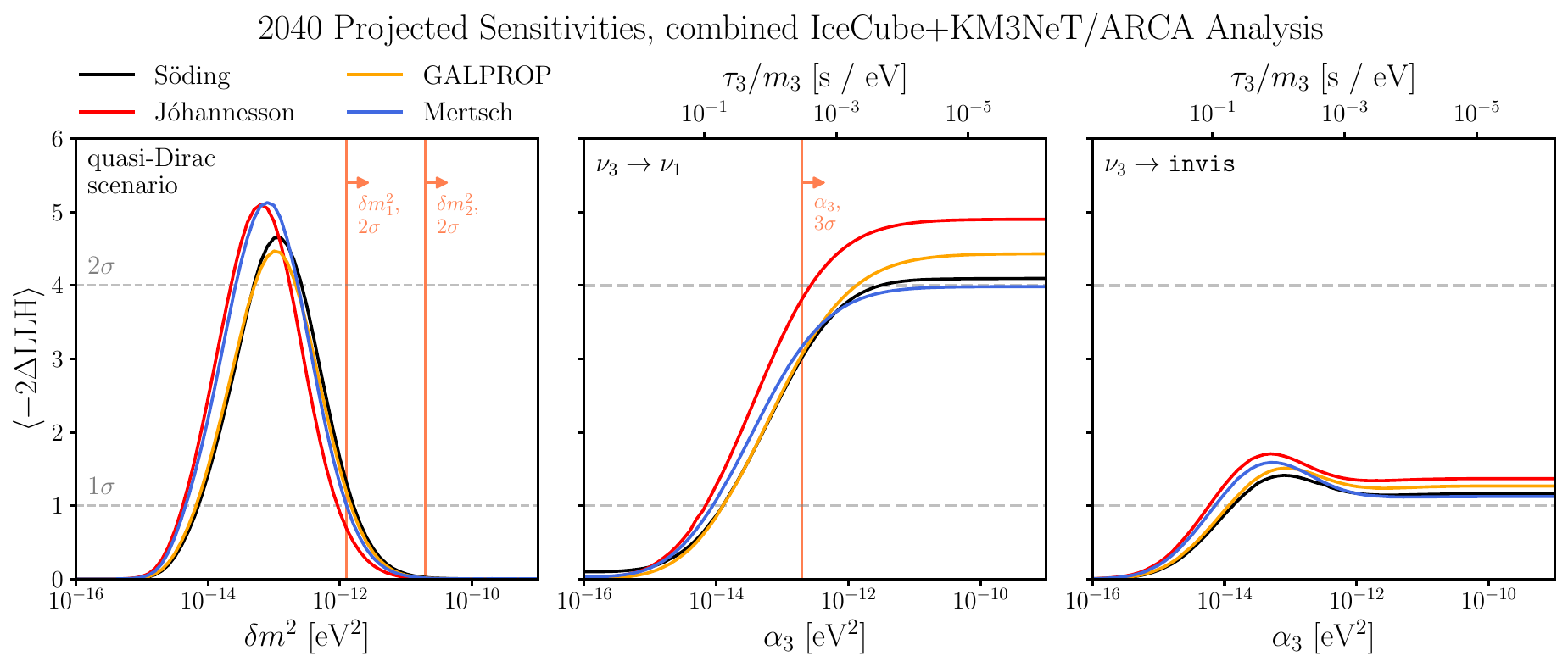}
  \caption[]{\textbf{\textit{Sensitivities to QD and neutrino decay models assuming different gas maps.}} We plot the sensitivities to a combined IceCube + KM3NeT 2040 analysis to the QD scenario (left), $\nu_3 \to \nu_1$ decay (middle), and $\nu_3 \to \mathtt{invis}$ decay (right) to four different gas map assumptions taken from Refs.~\cite{Soding:2025scd} (black),~\cite{Johannesson:2018bit} (maroon),~\cite{Fermi-LAT:2012edv} (orange), and \cite{Mertsch:2021gco,Mertsch:2023ghi} (blue). We indicate the same solar bounds shown in \Cref{fig:sensitivities}.} 
  \label{fig:sensitivities_diffgasmaps}
\end{figure*}

\begin{figure*}
  \includegraphics[width=\textwidth]{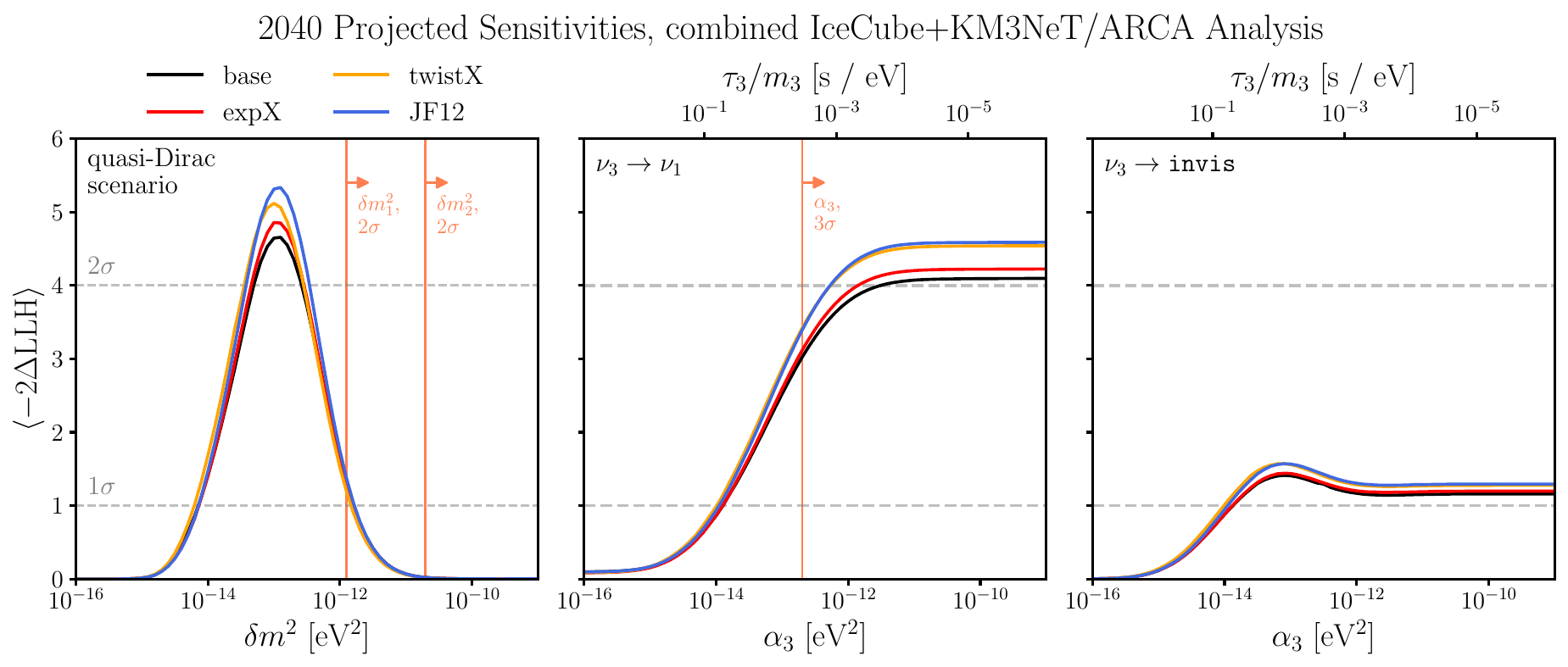}
  \caption[]{\textbf{\textit{Sensitivities to QD and neutrino decay models assuming different Galactic $B$-field models.}} We plot the sensitivities to a combined IceCube + KM3NeT 2040 analysis to the QD scenario (left), $\nu_3 \to \nu_1$ decay (middle), and $\nu_3 \to \mathtt{invis}$ decay (right) to four different $B$-field models taken from Refs.~\cite{Unger:2024cmf} and \cite{Jansson:2012rt}. We indicate the same solar bounds shown in \Cref{fig:sensitivities}.} 
  \label{fig:sensitivities_diffCRmaps}
\end{figure*}

Consequently, we find that our sensitivities to do not vary greatly between different models, as shown in Fig.~\ref{fig:sensitivities_diffgasmaps}. We similarly explore how the sensitivity is affected by varying the Galactic magnetic field model, which impacts the CR distribution, in Fig.~\ref{fig:sensitivities_diffCRmaps}. We find that the sensitivities are similarly robust.

\end{document}